\documentclass[11pt,a4paper]{article}
\usepackage{jcappub} 
\usepackage{bm}
\usepackage{amsmath}
\usepackage{mathrsfs}
\usepackage[mathscr]{euscript}
\usepackage{feynmp}
\usepackage{comment}
\usepackage{natbib}
\usepackage{ulem}

\allowdisplaybreaks[3]

\def\Mpl{M_{\rm Pl}}

\def\0{{(0)}}
\def\sig0{\dot{\sigma}_0}

\def\ph0{\dot{\phi}_0}

\title{
Peaky Production of \\[0.3ex] Light Dark Photon Dark Matter
}
\author[a]{Yuichiro Nakai,}
\author[b]{Ryo Namba,}
\author[c,d]{Ippei Obata}

\affiliation[a]{Tsung-Dao Lee Institute and School of Physics and Astronomy,\\
Shanghai Jiao Tong University, 800 Dongchuan Road, Shanghai 200240, China}
\affiliation[b]{RIKEN Interdisciplinary Theoretical and Mathematical Sciences (iTHEMS), \\
Wako, Saitama 351-0198, Japan}
\affiliation[c]{Max-Planck-Institut f{\"u}r Astrophysik, Karl-Schwarzschild-Str. 1, 85748 Garching, Germany}
\affiliation[d]{Kavli Institute for the Physics and Mathematics of the Universe (Kavli IPMU, WPI),UTIAS, The University of Tokyo, 5-1-5 Kashiwanoha, Kashiwa, Chiba, 277-8583, Japan}

\emailAdd{ynakai@sjtu.edu.cn}
\emailAdd{ryo.namba@riken.jp}
\emailAdd{ippei.obata@ipmu.jp}

\abstract{
We explore a mechanism to produce a light dark photon dark matter through a coupling between
the dark photon field and a spectator scalar field which plays no role in
the inflationary expansion of the Universe while rolling down its potential during the inflation.
The motion of the spectator field efficiently produces dark photons with large wavelengths
which become non-relativistic before the time of matter-radiation equality.
The spectrum of the wavelengths is peaky
so that the constraint from the isocurvature perturbation can be evaded.
The correct relic abundance is then achieved over a wide range of the dark photon mass down to $10^{-13} \ \text{eV}$.
Our mechanism favors high-scale inflation models which can be tested in future observations.
Furthermore, fluctuations of the dark photon field during inflation produce gravitational waves
detectable at future space-based interferometers and/or pulsar timing array experiments.
}

\begin{document}

\maketitle

\section{Introduction}

Ultra-light dark matter (DM) can in principle account for the issues around the structures of small scales (see \cite{Ferreira:2020fam} for a review).
A light spin-1 vector boson is an intriguing candidate of DM in our Universe.
Such a dark photon DM can have a wide range of mass, extending down to the fuzzy DM region
\cite{Hu:2000ke,Hui:2016ltb}.
Accordingly, searches for the dark photon DM have been conducted from various directions
(see refs.~\cite{Essig:2013lka,Antypas:2022asj} for reviews).
The Lyman-$\alpha$ constraint puts a lower bound on the mass, $m_{\gamma'} \gtrsim 10^{-20} \, \rm eV$
\cite{Irsic:2017yje}.
The black hole superradiance constraint spreads over a range of the dark photon mass,
$10^{-20} \, {\rm eV} \lesssim m_{\gamma'} \lesssim 10^{-11} \, \rm eV$
\cite{Arvanitaki:2009fg,Arvanitaki:2010sy,Baryakhtar:2017ngi,Cardoso:2017kgn,Cardoso:2018tly}.
For a light dark photon DM kinetically mixing with the photon,
additional constraints such as galactic heating~\cite{Dubovsky:2015cca}
and distortions of the cosmic microwave background (CMB)~\cite{Arias:2012az}
give upper bounds on the size of the kinetic mixing.
Moreover, a number of experiments are being programmed for further explorations.
For instance, DM Radio~\cite{Chaudhuri:2014dla,Silva-Feaver:2016qhh}
will search for the dark photon DM with a range of mass,
$10^{-12} \, {\rm eV} \lesssim m_{\gamma'} \lesssim 10^{-3} \, \rm eV$.
Recently, it has been revealed that laser interferometers for the detection of gravitational waves (GWs)
are also sensitive to the vector DM with mass $10^{-18} \, {\rm eV} \lesssim m_{\gamma'} \lesssim 10^{-11} \, \rm eV$ \cite{Michimura:2020vxn,Morisaki:2020gui,Nakatsuka:2022gaf}.
Therefore the phenomenology of a new vector boson DM is an exciting arena for the DM searches.

Depending on the dark photon mass $m_{\gamma'}$,
different mechanisms for the production of the dark photon DM have been proposed.
Inflationary fluctuations can give the correct DM abundance for $m_{\gamma'} \gtrsim \mu \rm eV$
\cite{Graham:2015rva}.
As in the case of axions, the dark photon DM can also be produced by a misalignment mechanism
\cite{Nelson:2011sf},
while a highly tuned coupling to the curvature is required
\cite{Arias:2012az}.
Refs.~\cite{Agrawal:2018vin,Dror:2018pdh,Co:2018lka} have realized a light dark photon DM
whose mass is below $\mu \rm eV$ by the oscillation of an axion-like field coupling to the dark photon
or a scalar field charged under the dark $U(1)$.
A network of cosmic strings can also produce a light dark photon DM
\cite{Long:2019lwl}.

A dark photon coupling to the inflaton provides another attractive possibility for the dark photon DM production.
Ref.~\cite{Bastero-Gil:2018uel} has introduced a coupling $\varphi F' \widetilde{F}'$
where $\varphi$ denotes the inflaton, $F'$ is the dark photon field strength and $\widetilde{F}'$ is its dual.
It was discussed that a dark photon DM with mass $m_{\gamma'} \gtrsim \mu \rm eV$ is produced through tachyonic instability during inflation.
A more generic coupling $I(\varphi)^2 F' F'$ with some function $I$ has been explored in refs.~\cite{Nakayama:2019rhg,Nakai:2020cfw}.
This type of coupling in a cosmological context was first introduced as a mechanism of inflationary magnetogenesis \cite{Ratra:1991bn} and has been actively studied \cite{Demozzi:2009fu,Barnaby:2012tk,Fujita:2012rb,Fujita:2013qxa,Fujita:2014sna,Ferreira:2013sqa,Ferreira:2014hma,Green:2015fss,BazrafshanMoghaddam:2017zgx,Kobayashi:2014sga,Fujita:2016qab,Kanno:2008gn,Watanabe:2009ct,Watanabe:2010fh,Gumrukcuoglu:2010yc,Namba:2012gg,Bartolo:2012sd,Naruko:2014bxa}. A crucial feature of the coupling is that, depending on the achieved time dependence of $I$, diverse shapes of the produced dark photon spectrum can be realized. Thanks to this nature, dark photons with large wavelengths can be efficiently produced by the inflaton motion.
The spectrum of produced dark photons has a peak around the wavelength that exits the horizon at the earliest time. 

The main target of the present paper is the production of a light dark photon DM with a range of mass,
$10^{-11} \, {\rm eV} \lesssim m_{\gamma'} \lesssim 10^{-6} \, \rm eV$,
which is above the mass region constrained by the black hole superradiance
and below the region in which inflationary fluctuations can give the correct relic abundance.
This range of mass is covered by the future DM radio experiment. 
Instead of a direct coupling to the inflaton, we here consider a coupling between
the dark photon field and a spectator scalar field which does not play a role in
the inflationary expansion but whose motion along its potential during inflation leads to production of the dark photon DM.
The motion of the spectator field can efficiently produce dark photons with large wavelengths
which become non-relativistic before the time of matter-radiation equality.
It has been known that the production of a light dark photon DM through a coupling to the inflaton
would receive a stringent constraint from the isocurvature perturbation
\cite{Nakayama:2020rka}.
On the other hand, in the present scenario, the wavelength spectrum of produced dark photons is peaky
so that the constraint from the isocurvature perturbation 
based on the CMB observations
can be evaded.
The correct relic abundance is then achieved over the target mass range and below, and we show that the mechanism favors high-scale inflation models to provide the lower dark photon mass.
In addition, fluctuations of the dark photon field during inflation produce GWs
which can be detected at future space-based interferometers such as DECIGO~\cite{Seto:2001qf}, BBO \cite{Crowder:2005nr} or $\mu$Ares \cite{Sesana:2019vho}
and also pulsar timing array experiments
such as SKA \cite{Kramer:2004hd}.

The rest of the paper is organized as follows.
In section~\ref{production}, we discuss the production of a light dark photon DM through a coupling between
the dark photon field and a spectator scalar field.
In Section \ref{energydensity}, we compute the power spectrum of the dark photon.
Section~\ref{DMabundance} then explores the parameter space that the correct DM abundance is realized.
In section~\ref{tensormodes}, the generation of tensor modes is investigated. 
Section~\ref{conclusion} is devoted to conclusions and discussions.

\section{A spectator model}
\label{production}

We consider a model of a light dark photon DM in which
a dark photon field $A'_\mu$ is kinetically coupled to a spectator scalar field $\sigma$.
The total action is given as follows:
\begin{align}
\label{eq:action}
S = \int d^4x\sqrt{-g}\left[\dfrac{\Mpl^2}{2}R + \mathcal{L}_{\rm inf} - \dfrac{1}{2}\partial_\mu\sigma\partial^\mu\sigma - V(\sigma) - \dfrac{I(\sigma)^2}{4}F'_{\mu\nu}F'^{\mu\nu} - \dfrac{1}{2}m_{\gamma'}^2A'_\mu A'^\mu \right] \ ,
\end{align}
where $R$ is the Ricci scalar 
associated with the $4$-D metric $g_{\mu\nu}$, whose determinant is denoted by $g$,
$F'_{\mu\nu} = \partial_\mu A'_\nu - \partial_\nu A'_\mu$ is the field strength of the dark photon, and $\mathcal{L}_{\rm inf}$ is the Lagrangian density of the inflaton.
In what follows, we do not specify the inflaton dynamics
except for the assumption that it can be well approximated by de Sitter for our purpose. In \eqref{eq:action},
$V(\sigma)$ is the potential of the spectator field $\sigma$, and $I(\sigma)$ is a function of $\sigma$
which will be specified later. 
The last term denotes a Proca term of the dark photon with a constant mass $m_{\gamma'}$.
We assume that there is no direct coupling between the dark photon and the Standard Model (SM) sector and no mixing with the ordinary photon.

A crucial point of our work is to turn on a nontrivial evolution of the dark photon production. This enables a non-monotonic spectrum of the produced dark photon, resulting in a spectral peak at a certain wavenumber $k_{\rm peak}$. This additional freedom is essential to evade the stringent constraint on the CMB isocurvature modes. The evolution of the production is controlled by the function $I(\sigma)$ in \eqref{eq:action}, whose behavior inherits the dynamics of $\sigma$. In the following subsections, we hence consider a slow but non-constant homogeneous motion of $\sigma$ and then perform the computation of the corresponding production of the dark photon.

\subsection{Background dynamics}

Let us first consider the background dynamics of the model.
Throughout the present work, we assume that there is no vacuum expectation value of the vector field:%
\footnote{This assumption in fact depends on the scales of interest. Even if the classical background of $A_\mu'$ is absent, the dark photon field can be generated from the quantum vacuum, as in our current consideration. If the produced field has a wavelength larger than a certain scale with a nonzero variance, then each local patch smaller than that stochastically experiences a coherent field of dark photon wave, which, strictly speaking, breaks spatial isotropy \cite{Bartolo:2012sd}. However, we are here interested in generation of the dark photon on scales smaller than the CMB ones, and therefore the vanishing $1$-point function is a relevant assumption, at least under the consideration of the isocurvature constraints which is the main observational obstacle for dark photon production of long wavelength modes.
}
\begin{equation}
\label{eq:homogeneousA}
\langle A'_\mu \rangle = 0 \ .
\end{equation}
Thanks to this assumption, 
and ensuring \textit{a posteriori} that the energy density of the dark photon is a subdominant component,
we can consider the flat
Friedmann-Lema\^{i}tre-Robertson-Walker (FLRW) metric,
\begin{equation}
\label{eq:FLRWmetric}
ds^2 = -dt^2 + a(t)^2d\bm{x}^2 = a(\tau)^2(-d\tau^2 + d\bm{x}^2) \ ,
\end{equation}
as the background spacetime.
Here, $a$ is the scale factor and
$\tau$ denotes the conformal time,
$d \tau = dt / a$.
Throughout the paper, we assume that inflation proceeds as de Sitter essentially, and the time dependence of the scale factor is well approximated by $a \simeq \exp\left( H t \right) \simeq - 1 / (H\tau)$, where $H \equiv \partial_t a / a$ is the Hubble rate during inflation.

For the kinetic function $I$ of $\sigma$, we adopt the following simple exponential form:
\begin{equation}
I(\sigma) = I_0\exp\left( \dfrac{\sigma}{\Lambda_1} \right) \, , \label{eq: I}
\end{equation}
with some dimension-$1$ energy scale $\Lambda_1$.
The normalization constant $I_0$ is assumed to be $I_0=1$ to realize $I(\sigma\rightarrow0) = 1$ after inflation ends when $\sigma$ settles into its potential minimum and recovers the canonical kinetic term of the dark photon field.
The crucial parameter that dictates production of the dark photon, especially of the transverse modes, is the rate of the time variation of $I$ with respect to the number of e-foldings $N = \ln a$, i.e.,
\begin{equation}
n \equiv \dfrac{dI/dN}{I} 
= \frac{d \sigma / d N}{\Lambda_1}
\ , \qquad dN = Hdt \, .
\label{eq: ndef}
\end{equation}
The critical values of $n$ for dark photon production are $\vert n \vert = 2$: for $\vert n \vert > 2$, the associated dark electric or magnetic field, or both, goes through exponential amplifications \cite{Bamba:2003av}.

With the form of $I$ fixed by \eqref{eq: I}, we take the freedom to choose the spectator potential $V(\sigma)$ so that the dark photon production occurs with nontrivial time dependence. In this regard, we consider a toy model,
\begin{equation}
V(\sigma) = \mu^4\tanh^2\left(\dfrac{\sigma}{\Lambda_2}\right) \ , \label{eq: tanh2}
\end{equation}
where $\mu$ is a potential energy scale and $\Lambda_2$ is another energy scale which characterizes the steepness of the potential.
This kind of functional form is sometimes considered in the context of the $\alpha$-attractor model \cite{Kallosh:2013yoa},
while in our current study we assume that another independent sector drives inflation.
To characterize the energy density of the spectator field $\sigma$, playing no role in the inflationary expansion,
we define the following ratio,
\begin{equation}
R \equiv \dfrac{\mu^4}{3\Mpl^2H^2} \ll 1 \ .
\end{equation}
The Hubble slow-roll parameter $\epsilon_H \equiv -\dot{H}/H^2$
(dot represents derivative with respect to the physical time $t$) is given by
\begin{equation}
\epsilon_H = \epsilon_\phi + \epsilon_\sigma + \epsilon_{\gamma'} \ll 1 \label{eq: slow1} \ , 
\end{equation}
where each contribution is defined as $\epsilon_\alpha \equiv \left( \rho_\alpha + p_\alpha \right) / (2 \Mpl^2 H^2)$ ($\alpha = \phi, \, \sigma, \, \gamma'$), where $\rho_\alpha$ and $p_\alpha$ are the corresponding energy density and pressure, respectively, and reads
\begin{align}
\epsilon_\phi & = \dfrac{\dot{\phi}^2}{2\Mpl^2H^2} \ , \quad \epsilon_\sigma = \dfrac{\dot{\sigma}^2}{2\Mpl^2H^2} \label{eq: slow2} \ , \\
\epsilon_{\gamma'} & = \dfrac{1}{\Mpl^2H^2} \left\langle \dfrac{1}{3a^2}I^2\left(F'_{0i} F'_{0i} + \dfrac{1}{2a^2}F'_{ij} F'_{ij}\right) + \dfrac{1}{2}m_{\gamma'}^2\left(A{'}_{0}^2 + \dfrac{1}{3a^2}A{'}_i A{'}_i\right) \right\rangle \ .
\label{eq:epsilon_gamma}
\end{align}
Here, $\epsilon_{\gamma'}$ denotes a contribution from the backreaction of
the dark photon field and we assume that its magnitude is small in comparison with the other contributions.
Note that, in \eqref{eq:epsilon_gamma}, the expression assumes the physical time (not conformal time), and the bracket $\langle \bullet \rangle$ denotes the vacuum average.
For our analytical convenience, we take the Hubble parameter as a constant value $H = \text{const}.$ to solve the dynamics of the spectator-dark photon system.

The equation of motion for the spectator field is
\begin{equation}
\dfrac{d^2\sigma}{dN^2} + 3\dfrac{d\sigma}{dN} + \dfrac{V_\sigma}{H^2} = -\dfrac{II_\sigma}{2H^2}\langle F'_{\mu\nu}F{'}^{\mu\nu} \rangle \ , \label{eq: sigmaeom}
\end{equation}
where the subscript $\sigma$ denotes derivative with respect to $\sigma$, and the time dependence of $H$ is assumed to be negligible ($\vert \partial_N H / H \vert \ll 3$).
The right-hand side describes a backreaction effect on the motion of the spectator field.
We consider the parameter space where this effect is negligible
as we will discuss later.
Then, assuming that the first term of the left-hand side
in Eq.~\eqref{eq: sigmaeom} is also negligible, we can use the slow-roll condition $3d\sigma/dN \simeq -V_\sigma/H^2$ and obtain from \eqref{eq: ndef}
\begin{equation}
n \simeq -\dfrac{V_\sigma}{3H^2\Lambda_1} = -2cR\dfrac{\Mpl^2}{\Lambda_1^2}\dfrac{\tanh(\tilde{\sigma})}{\cosh^2(\tilde{\sigma})} \ . \label{eq: na}
\end{equation}
Here, we have defined dimensionless quantities, $c \equiv \Lambda_1/\Lambda_2$ and $\tilde{\sigma} \equiv \sigma/\Lambda_2$.
Regarding the slow-roll solution of $\sigma$, we find that $\tilde{\sigma}$ and $n$ are described in terms of the Lambert W-function as
\begin{align}
\tilde{\sigma}(N) &= 
\text{arcsinh} \left[ \sqrt{W(y)} \right] \ ,
\label{eq:sigma_sol} \\
n(N) &= -2cR \, \dfrac{\Mpl^2}{\Lambda_1^2} \, 
\dfrac{\sqrt{W(y)}}{\left[ 1 + W(y) \right]^{3/2}} \ ,
\label{eq: n} \\
y & \equiv - 4 R \, \frac{\Mpl^2}{\Lambda_2^2} \, N + 2\ln\left[\sinh(\tilde{\sigma}_i)\right] + \sinh^2(\tilde{\sigma}_i) \ ,
\label{eq: y_by_sigma}
\end{align}
where $\tilde{\sigma}_i = \tilde{\sigma}(0) > 0$ is an initial value of $\tilde{\sigma}$.
Inversely, using the definition of the W-function $W(z e^z) = z$, the number of e-foldings is written as
\begin{equation}
N = \dfrac{\Lambda_2^2}{4R\Mpl^2}\left[ 2\ln\left[\dfrac{\sinh(\tilde{\sigma}_i)}{\sinh(\tilde{\sigma})}\right] + \sinh^2(\tilde{\sigma}_i) - \sinh^2(\tilde{\sigma}) \right] \ .
\end{equation}
Note that, since we are requiring the weak coupling of the dark photon to the hidden sector, i.e.~$I > 1$, during inflation, we choose the branch $\sigma > 0$, provided $\Lambda_{1,2} > 0$.

%
\begin{figure}[tbp]
\center
  \includegraphics[width=0.5\linewidth]{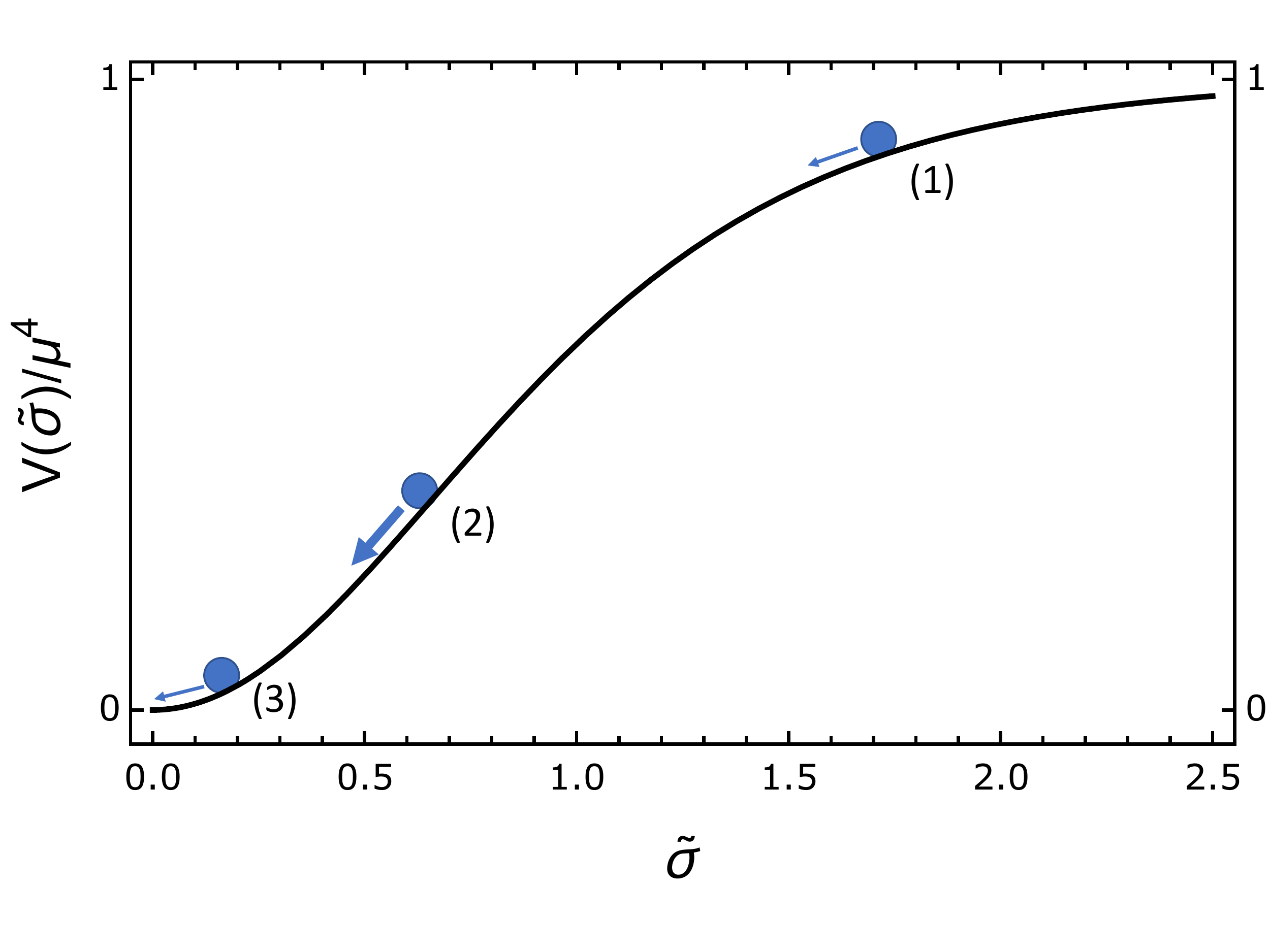}
  \vspace{-0.3cm}
  \caption{
  A schematic dynamics of the spectator field on the $\tanh^2$ potential \eqref{eq: tanh2} in our scenario.
  (1) The initial value is located
  toward the top of the hill and the velocity monotonically increases.
  (2) At a certain time, the spectator field crosses an inflection point of the potential and the velocity gets maximized.
  (3) The velocity turns to decrease toward the minimum of the potential until inflation ends. 
  }
 \label{fig:ArcTan}
\end{figure}
%

We are interested in a situation where the magnitude of $|n|$ is initially small but becomes greater than 2 at some point during inflation.
Namely, the dark photon production occurs at the intermediate stage of inflation, and its overproduction on large scales is avoidable.
To realize such a situation, we consider the following scenario: the spectator field rolls down on the potential and its velocity experiences a peak value at an intermediate time of inflation (see Figure~\ref{fig:ArcTan}).
At an initial stage, $\tilde{\sigma}$ is near the top of the potential and $|n| < 2$ is satisfied.
Then, the velocity of the scalar field gradually increases and gets maximized when it goes through an inflection point $\tilde{\sigma} = \tilde{\sigma}_{\rm IP}$ on the potential,
\begin{equation}
V_{\sigma\sigma}(\tilde{\sigma}_{\rm IP}) = 2 \, \dfrac{\mu^4}{\Lambda_2^2} \,
\dfrac{1-2\sinh^2(\tilde{\sigma}_{\rm IP})}{\cosh^4(\tilde{\sigma}_{\rm IP})}
= 0 \qquad \longleftrightarrow \qquad \tilde{\sigma}_{\rm IP} \simeq 0.658 \, ,
\label{eq:Vsigsig}
\end{equation}
at a certain number of e-foldings $N = N_{\rm IP}$, assuming $n(N_{\rm IP}) < -2$.
After that, the velocity decreases and finally the scalar field reaches
the bottom of the potential.
We define the number of e-foldings $N_1$ and $N_2$ at which $n(N_1) = n(N_2) = -2$ and $N_1 < N_{\rm IP} < N_2$ are satisfied.
Using Eq.~\eqref{eq: na}, they are obtained as
\begin{equation}
N_p = \dfrac{1}{4c}\left[\dfrac{\tanh(\tilde{\sigma}_{p})}{\cosh^2(\tilde{\sigma}_{p})}\left( 2\ln\left[\dfrac{\sinh(\tilde{\sigma}_i)}{\sinh(\tilde{\sigma}_{p})}\right] + \sinh^2(\tilde{\sigma}_i) - \sinh^2(\tilde{\sigma}_{p}) \right) \right] \quad (p=1,2) \ ,
\end{equation}
where $\tilde{\sigma}_{p} = \tilde{\sigma}(N_{p})$.
Therefore, the parameter $c$, characterizing the slope of the potential, 
partially determines a duration of e-folds for which the particle production occurs. 
To obtain $N_p = \mathcal{O}(10)$, we may choose $c = \mathcal{O}(10^{-2})$ for $\tilde{\sigma}_i = \mathcal{O}(1)$.

\begin{figure}[!t]
 \begin{minipage}{0.5\hsize}
  \begin{center}
   \includegraphics[width=70mm]{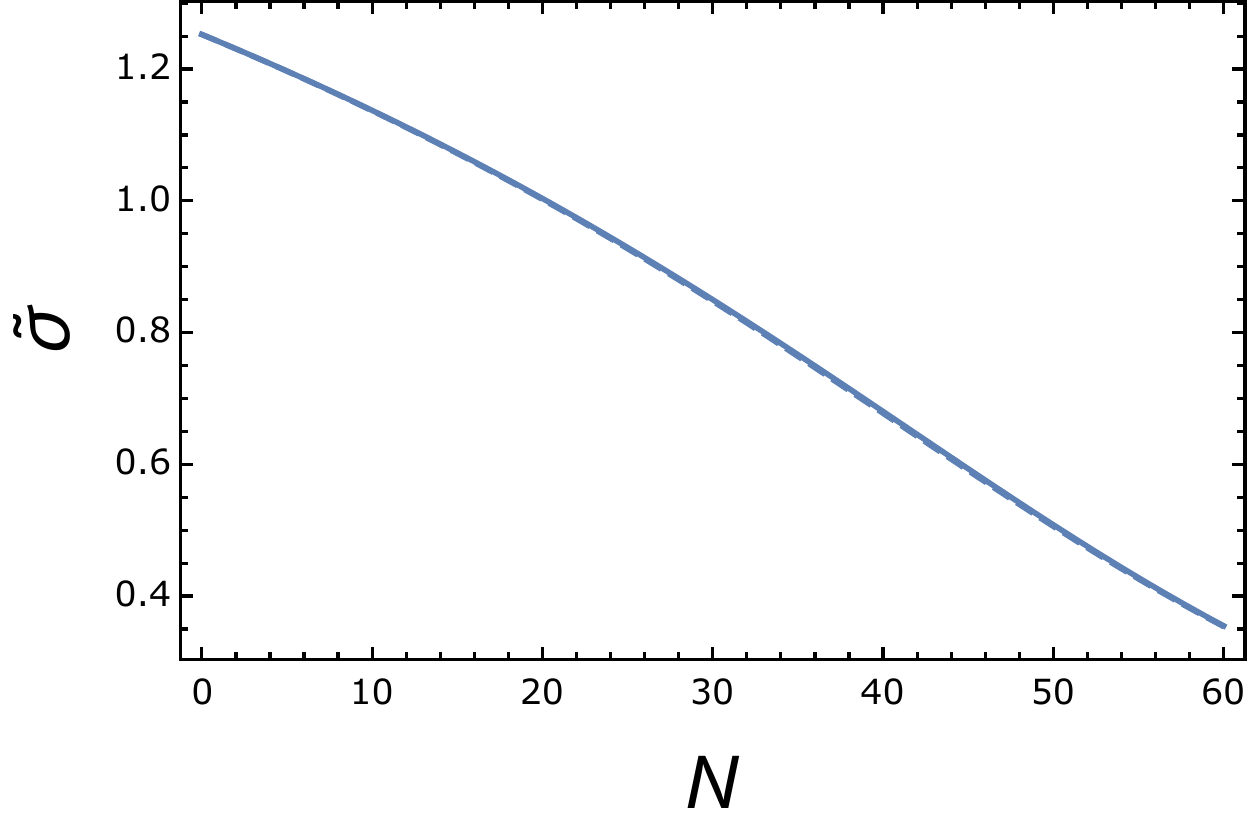}
  \end{center}
 \end{minipage}
 \begin{minipage}{0.5\hsize}
  \begin{center}
   \includegraphics[width=70mm]{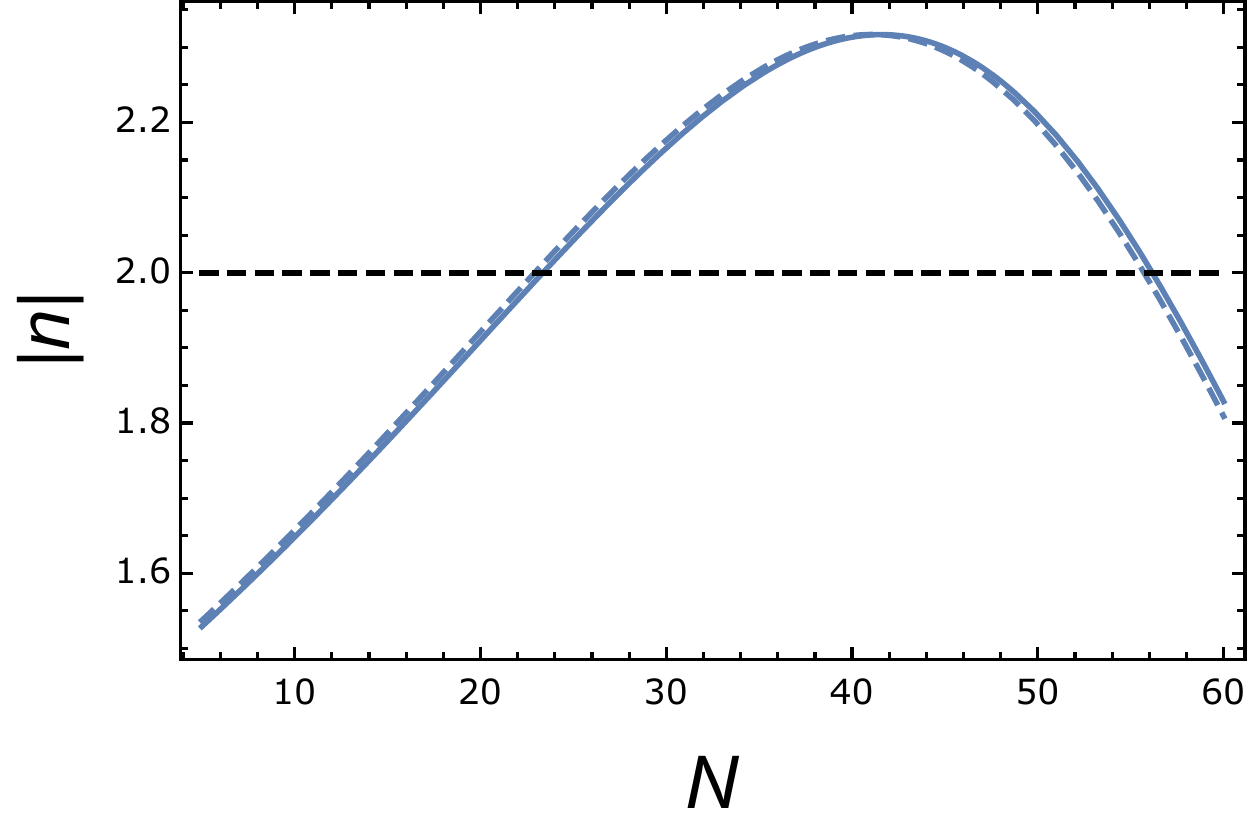}
  \end{center}
 \end{minipage}
   \caption{The time evolution of $\tilde{\sigma}$ (left panel) and $n$ (right panel). The solid and dashed lines represent the numerical and slow-roll approximate solutions, respectively.
   Here, we take $R = 10^{-2}$, $\tilde{\sigma}_i = 1.2525$, $\Lambda_1 = 5\times10^{-3}\Mpl, \ c = 7.525\times10^{-3}$.
   }
  \label{fig:two}
\end{figure}

Figure~\ref{fig:two} shows the time evolution of $\tilde{\sigma}$ and
that of the corresponding $n$.
We set the inflation end at $N_{\rm end} = 60$.
The solid and dashed curves represent the numerical and slow-roll approximate solutions \eqref{eq:sigma_sol} and \eqref{eq: n}, respectively.
One can find that the slow-roll solution is well-fitted to the exact one.
At an initial stage of inflation, $\vert n \vert$ is smaller than 2 and
the vector field is not amplified.
At a certain time, $\vert n \vert$ becomes greater than 2
and the vector field starts to be amplified.
In the next section, we will show that the dark photon production
occurs at an intermediate stage of inflation.
As a result, the sourced power spectrum of the dark photon exhibits a peaky structure on small scales, reflecting the peaky evolution of $n$.

\subsection{Evolution of the dark photon field}

Let us analyze the evolution of the dark photon field kinetically coupled to the scalar field $\sigma$.
Since the vector field possesses a nonzero mass, its spatial components are decomposed into transverse modes and a longitudinal mode:
\begin{equation}
A'_i = A'^T_i + \partial_i\chi \ , \ \qquad \partial_iA'^T_i = 0 \ .
\end{equation}
Integrating out the non-dynamical component $A'_0$
by
\begin{align}
    A_0' = \frac{-I^2 \partial^2}{-I^2 \partial^2 + a^2 m_{\gamma'}^2} \, \partial_\tau \chi \; ,
    \label{eq:A0}
\end{align}
the quadratic action of the transverse modes and that of the longitudinal mode are respectively obtained as
\begin{align}
S_T &= \dfrac{1}{2}\int d\tau \, d\bm{x} \Big[I^2(\partial_\tau A'^T_i\partial_\tau A'^T_i - \partial_i A'^T_j\partial_i A'^T_j) - a^2m_{\gamma'}^2A'^T_iA'^T_i \Big] \ , \label{eq: transaction} \\
S_L &= \dfrac{1}{2}\int d\tau \, d\bm{x} \, a^2m_{\gamma'}^2\left[\partial_\tau\chi\left(\dfrac{-I^2\partial^2}{-I^2\partial^2 + a^2m_{\gamma'}^2}\partial_\tau\chi\right) - \partial_i\chi\partial_i\chi \right] \label{eq: longaction} \ ,
\end{align}
with $\partial^2 \equiv \partial_i\partial_i$.
Modulations of the kinetic couplings in
these actions are crucial for the evolution of the vector field.
To see this, we decompose the transverse modes ($2$ degrees of freedom) and the longitudinal one ($1$ d.o.f.)
in terms of their Fourier modes,
\begin{align}
IA'^T_i(\tau, \bm{x}) &= \int\dfrac{d\bm{k}}{(2\pi)^3}\left(\hat{V}^X_{\bm{k}}(\tau)e^X_i(\hat{\bm{k}}) + i\hat{V}^Y_{\bm{k}}(\tau)e^Y_i(\hat{\bm{k}})\right)e^{i\bm{k}\cdot\bm{x}} \ , 
\label{eq:fourier_AT} \\
\chi(\tau, \bm{x}) &= \int\dfrac{d\bm{k}}{(2\pi)^3}\dfrac{\hat{X}_{\bm{k}}(\tau)}{z_k}e^{i\bm{k}\cdot\bm{x}} \ , \qquad z_k \equiv \dfrac{a m_{\gamma'}I k}{\sqrt{I^2k^2 + a^2m_{\gamma'}^2}} \ ,
\label{eq:fourier_chi}
\end{align}
where $k \equiv \vert \bm{k} \vert$, superscripts $X$ and $Y$ label the polarization states, and the orthonormal polarization vectors take the forms
\begin{equation}
e^X_i(\hat{\bm{k}}) = (\cos\theta\cos\phi, \ \cos\theta\sin\phi, \ -\sin\theta) \ , \quad e^Y_i(\hat{\bm{k}}) = (-\sin\phi, \ \cos\phi, \ 0) \ ,
\end{equation}
with the unit wave vector $\hat{\bm{k}} = (\sin\theta\cos\phi, \ \sin\theta\sin\phi, \ \cos\theta)$.
They obey the properties:
$k_i e^\sigma_i(\hat{\bm{k}}) = 0
\ (\sigma = X,Y) , \ e^X_i(\hat{\bm{k}})e^X_i(\hat{\bm{k}}) = e^Y_i(\hat{\bm{k}})e^Y_i(\hat{\bm{k}}) = 1 , \ e^X_i(\hat{\bm{k}})e^Y_i(\hat{\bm{k}}) = 0$,
$e_i^{\sigma \, *}(\hat{\bm{k}}) = e_i^\sigma(\hat{\bm{k}}), \ e_i^X( - \hat{\bm{k}}) = e_i^X(\hat{\bm{k}})$, $e_i^Y( - \hat{\bm{k}}) = - e_i^Y(\hat{\bm{k}})$.%
\footnote{Note that we define the parity change $\hat{k} \to - \hat{k}$ by the simultaneous operations $\theta \to \pi - \theta$ and $\phi \to \phi + \pi$.}
The canonical variables $\hat{V}^{\sigma}_{\bm{k}}$ and $\hat{X}_{\bm{k}}$ satisfy the reality conditions,
\begin{equation}
    \hat{V}_{\bm{k}}^{X \, \dagger} = \hat{V}_{-\bm{k}}^X \; , 
    \qquad
    \hat{V}_{\bm{k}}^{Y \, \dagger} = \hat{V}_{- \bm{k}}^Y \; ,
    \qquad
    \hat{X}_{\bm{k}}^\dagger = \hat{X}_{-\bm{k}} \; .
\end{equation}
Note that we put $i$ in front of the $\hat{V}^Y_{\bm{k}}$ term in \eqref{eq:fourier_AT} so that $\hat{V}^Y_{\bm{k}}$ respects the above property of Hermitian conjugate.
In terms of these variables, the actions \eqref{eq: transaction} and \eqref{eq: longaction} are rewritten as
\begin{align}
S_T &= \dfrac{1}{2}\sum_{\sigma = X, Y}\int d\tau \, \dfrac{d\bm{k}}{(2\pi)^3}\left[ \partial_\tau\hat{V}^{\sigma\dagger}_{\bm{k}}\partial_\tau\hat{V}^{\sigma}_{\bm{k}} - \left( k^2 - \dfrac{\partial_\tau^2I}{I} + \dfrac{a^2m_{\gamma'}^2}{I^2} \right)\hat{V}^{\sigma\dagger}_{\bm{k}}\hat{V}^{\sigma}_{\bm{k}} \right] \ , \\
S_L &= \dfrac{1}{2}\int d\tau \, \dfrac{d\bm{k}}{(2\pi)^3}\left[ \partial_\tau\hat{X}^{\dagger}_{\bm{k}}\partial_\tau\hat{X}_{\bm{k}} - \left( k^2 - \dfrac{\partial_\tau^2 z_k}{z_k} + \dfrac{a^2m_{\gamma'}^2}{I^2}
\right)\hat{X}^{\dagger}_{\bm{k}}\hat{X}_{\bm{k}} \right] \ .
\end{align}
Then, decomposing $\hat{V}^{\sigma}_{\bm{k}}$, $\hat{X}_{\bm{k}}$
into creation/annihilation operators,
\begin{align}
\hat{V}^\sigma_{\bm{k}} &= V^\sigma_k \hat{a}^\sigma_{\bm{k}} + V^{\sigma*}_k \hat{a}^{\sigma\dagger}_{-\bm{k}} \ , \qquad [\hat{a}^\sigma_{\bm{k}}, \hat{a}^{\sigma'\dagger}_{-\bm{k'}}] = (2\pi)^3\delta^{\sigma\sigma'}\delta(\bm{k} + \bm{k'}) \ , \\
\hat{X}_{\bm{k}} &= X_k \hat{b}_{\bm{k}} + X^{*}_k \hat{b}^{\dagger}_{-\bm{k}} \ , \qquad [\hat{b}_{\bm{k}}, \hat{b}^{\dagger}_{-\bm{k'}}] = (2\pi)^3\delta(\bm{k} + \bm{k'}) \ ,
\end{align}
with the vacuum $\vert 0 \rangle$ defined by $\hat{a}_{\bm{k}}^\sigma \vert 0 \rangle = \hat{b}_{\bm{k}} \vert 0 \rangle = 0$,
we obtain the equations of motion for the mode functions as
\begin{align}
&\partial_\tau^2V_k + \left( k^2 - \dfrac{\partial_\tau^2I}{I} + \dfrac{a^2m_{\gamma'}^2}{I^2} \right)V_k = 0 \ , \label{eq: Ve1} \\
&\partial_\tau^2X_k + \left( k^2 - \dfrac{\partial_\tau^2z_k}{z_k} + \dfrac{a^2m_{\gamma'}^2}{I^2} \right)X_k = 0 \ , \label{eq: Xe1}
\end{align}
where the index $\sigma \, (=X,Y)$ of the transverse mode functions has been omitted since the two modes obey the same equation.
Historically, for analytical convenience, many of the studies on inflationary kinetic coupling models have assumed a special functional form of $I$ to make $n$ constant in time.
Such a case, with a negligible mass term, is briefly summarized in Appendix \ref{app:constantn}.
In the case of small mass, the longitudinal mode obeys the equation of motion as in \eqref{eq:eom_long_largep} and does not experience an exponential enhancement, in contrast to the transverse modes when $\vert n \vert > 2$ is realized. Thus the contribution to the dark photon production from the longitudinal mode is always subdominant in the parameter space of our interest and shall be disregarded in the following consideration.

In order to find the solution of the transverse modes with a mild time dependence of $n(\tau)$, we assume a negligible mass of the dark photon, and then
Eq.~\eqref{eq: Ve1} is rewritten as
\begin{equation}
\partial_\tau^2V_k + \left( k^2 - \dfrac{n(\tau)(n(\tau)+1)+dn/dN}{\tau^2}  \right)V_k = 0 \ , \label{eq: Ve3}
\end{equation}
with a constant $H$.
Neglecting a small correction of the velocity term $dn/dN$,
we obtain a differential equation of the same form as Eq.~\eqref{eq: Ve2}.
Although in the case of time-dependent $n(\tau)$ there would be no analytically closed form of the solution to eq.~\eqref{eq: Ve3} in general, there exists a useful technique called uniform approximation \cite{Habib:2002yi,Habib:2004kc,Lorenz:2008et}, which we employ here. By introducing variables
\begin{equation}
g(\tau) \equiv \dfrac{\nu^2(\tau)}{\tau^2} - k^2 \ , \qquad 
f(\tau) \equiv 
\begin{cases}
\displaystyle
- \left[ \dfrac{3}{2}\int_\tau^{\tau_*} d\tilde{\tau}\sqrt{- g(\tilde{\tau})} \right]^{2/3} \; , & \quad \tau < \tau_* \; , \vspace{1mm} \\
\displaystyle
\left[ \dfrac{3}{2}\int_{\tau_*}^\tau d\tilde{\tau}\sqrt{g(\tilde{\tau})} \right]^{2/3} \; , & \quad \tau_* < \tau \; ,
\end{cases}
\label{eq: UA}
\end{equation}
with $\nu^2 \equiv (1+4n(n+1))/4$
and the turning point $\tau_* < 0$ defined by $g(\tau_*) = 0$,
the solution of Eq.~\eqref{eq: Ve3} is well described by the following linear combination of Airy functions:
\begin{equation}
V^{\rm UA}_k \equiv A_k\left(\dfrac{f}{g}\right)^{1/4}\text{Ai}(f) + B_k\left(\dfrac{f}{g}\right)^{1/4}\text{Bi}(f) \ ,
\label{eq:V_UA}
\end{equation}
where the superscript ``UA'' is to remind (the leading order of) the uniform approximation.
To utilize $V_k^{\rm UA}$ as an approximate solution to \eqref{eq: Ve3},
the coefficients $A_k$ and $B_k$ are chosen to realize the adiabatic initial condition in the sub-horizon regime and found to be
\begin{equation}
A_k = iB_k \ , \qquad B_k = \sqrt{\dfrac{\pi}{2}}e^{i\theta} \ ,
\end{equation}
where $\theta$ is an overall phase factor irrelevant in the following discussion.
Then, using the asymptotic behavior of Airy functions, the solution on the super-horizon regime is approximately given by
\begin{align}
V^{\rm UA}_k & \simeq \dfrac{B_k}{\sqrt{\pi}g^{1/4}}\exp\left(\dfrac{2}{3}f^{3/2}\right) 
\simeq e^{i\theta}\dfrac{\sqrt{-k\tau}}{\sqrt{2\nu k}} \exp \left( \int^\tau_{\tau_*} d \tilde\tau \sqrt{\frac{\nu^2(\tilde\tau)}{\tilde\tau^2} - k^2} \right)
\label{eq:V_UA_approx}
\end{align}
for $- k \tau \ll \vert \nu(\tau) \vert$.
One caution to note is that this leading-order solution of the uniform approximation captures the correct spectral behavior, but its amplitude can be slightly off. Even in the case of constant $n$, or equivalently constant $\nu$, denoted by $\nu_0$, the comparison to the known exact solution using the Hankel function (see e.g.~\cite{Barnaby:2012tk}), $V_k^{\rm exact}$, reveals the difference
\begin{align}
V_k^{\rm UA} \simeq \frac{e^{i \theta}}{\sqrt{k}} \, e^{-\nu_0} \left( \frac{2\nu_0}{- k \tau} \right)^{\nu_0 - \frac{1}{2}} \; , \quad
V_k^{\rm exact} \simeq \frac{e^{i \theta}}{\sqrt{k}} \, \frac{2^{\nu_0 - 1} \Gamma(\nu_0)}{\sqrt{\pi} \left( - k \tau \right)^{\nu_0 - \frac{1}{2}}}  \; , \quad
\bigg\vert \frac{V_k^{\rm UA}}{V_k^{\rm exact}} \bigg\vert \simeq \frac{\sqrt{2\pi} \, \nu_0^{\nu_0 - \frac{1}{2}}}{e^{\nu_0} \Gamma(\nu_0)}
\simeq 0.95 \; ,
\end{align}
where the last numerical value is evaluated for the case $n_0 = -2$, or $\nu_0 = 3/2$.
This is due to the error of the truncation at the leading order. The uniform approximation can accommodate the calculations of higher-order terms in an iterative manner for an arbitrary $\nu(\tau)$ \cite{olver1974asymptotics}. For constant $\nu = \nu_0$, for example, the first sub-leading correction gives $V_k^{\rm UA (2)} = V_k^{\rm UA} / (12 \nu_0)$ \cite{Habib:2004kc}, for which the difference from the exact result becomes $\vert (V_k^{\rm UA} + V_k^{\rm UA (2)}) / V_k^{\rm exact} \vert \simeq 0.9999$ for $\nu_0 = 3/2$. Nonetheless, our leading-order solution \eqref{eq:V_UA}, and its super-horizon limit \eqref{eq:V_UA_approx}, gather the correct time evolution and the spectral feature, as can be seen in Fig.~\ref{fig:IA}. We thus use the leading order and admit the $\sim 5 \, \%$ error in the amplitude in the following considerations.

\section{The energy density spectrum}
\label{energydensity}

Based on the nontrivial background dynamics and the resulting characteristic production of dark photons described in the previous sections, we compute their energy density and show that its power spectrum can be peaked at scales much smaller than those of CMB observations but with wavelengths large enough to realize small-mass cold dark matter.
The energy density $\rho_{\gamma'}$ is the (0,0)-component of the energy-momentum tensor of the dark photon $T^{A'}_{\mu\nu} = -2\delta S[A']/\delta g^{\mu\nu}$,
which is calculated as
\begin{equation}
\rho_{\gamma'} = -T^{A'0}{_0} = \dfrac{1}{2a^4}\left[ I^2\left(F'_{0i} F'_{0i} + \dfrac{1}{2}F'_{ij} F'_{ij}\right) + a^2m_{\gamma'}^2\left(A{'}_{0}^2 + A{'}_i A{'}_i\right) \right] \ ,
\end{equation}
where this expression is compatible with the conformal time (not the physical one).
The vacuum averaged energy density is split into two parts:%
\footnote{The energy density is quadratic in the field $A_\mu$, and there in principle exist cross terms between the transverse and longitudinal modes. They do not contribute to the vacuum average for two reasons: the first is that the different polarization modes are decoupled from each other at the linear order, and thus their cross correlations vanish. The second is that the cross terms always appear as total (spatial) derivatives, which is a consequence of the background isotropy and homogeneity, together with a vanishing vector vev $\langle A_\mu \rangle =0$, and therefore vanish when vacuum average is taken.}
%
\begin{figure}[tbp]
\center
\begin{tabular}{ccc}
\begin{minipage}[t]{0.3\hsize}
  \includegraphics[width=50mm]{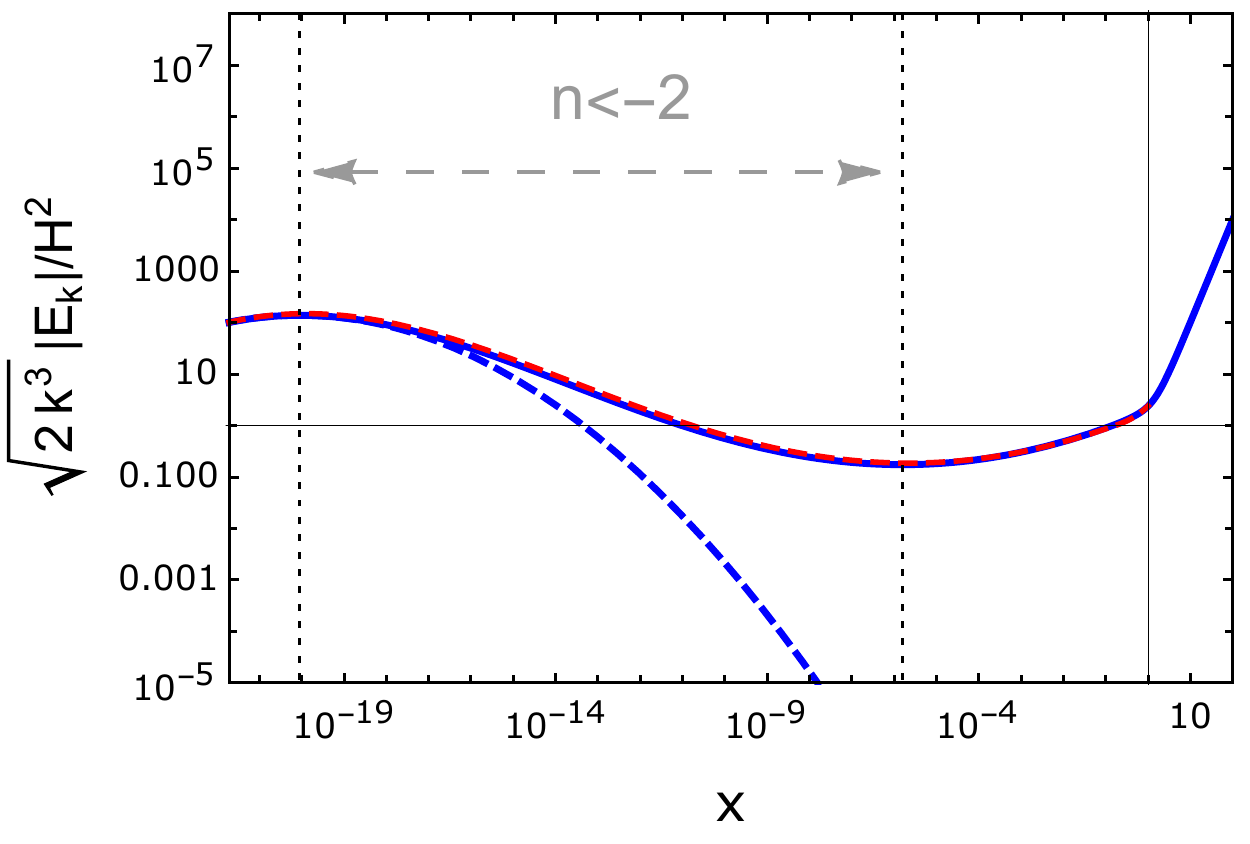}
  \end{minipage} &
  \begin{minipage}[t]{0.3\hsize}
  \includegraphics[width=50mm]{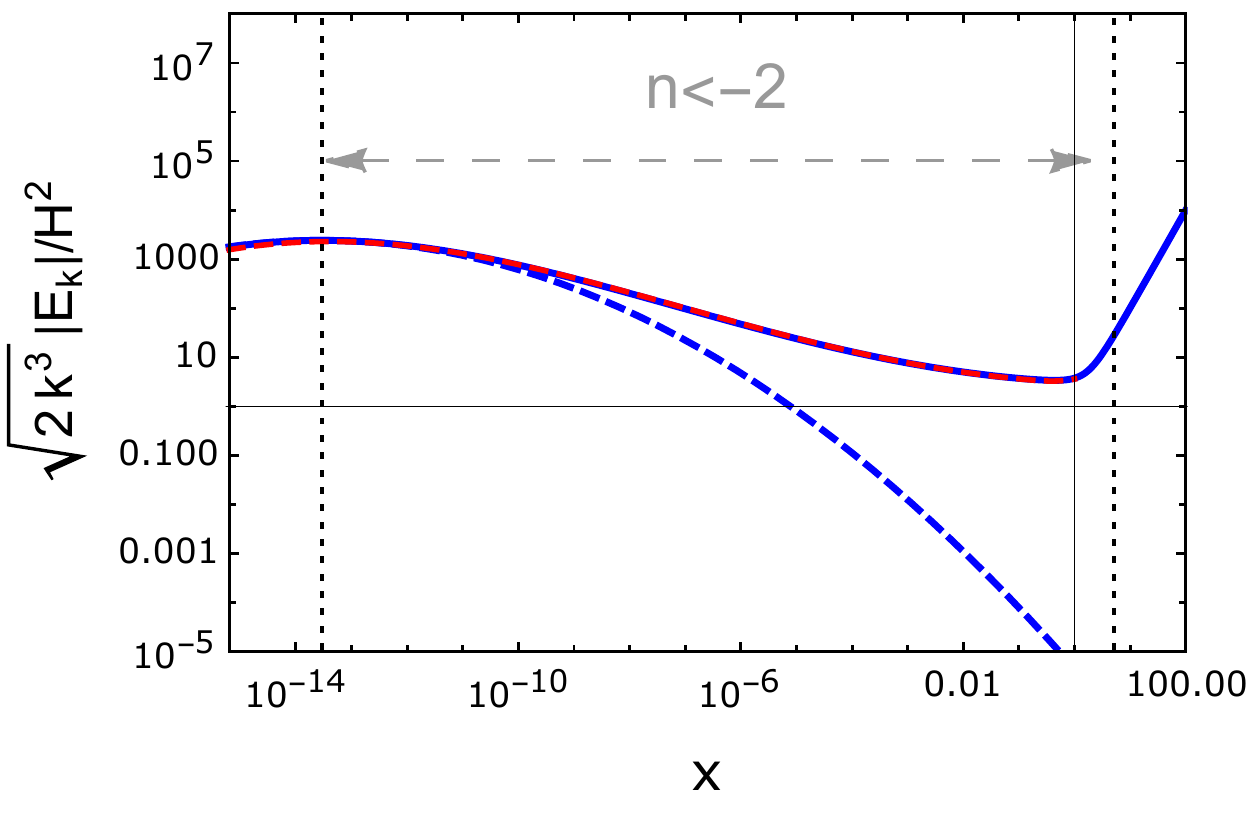}
  \end{minipage} &
  \begin{minipage}[t]{0.3\hsize}
  \includegraphics[width=50mm]{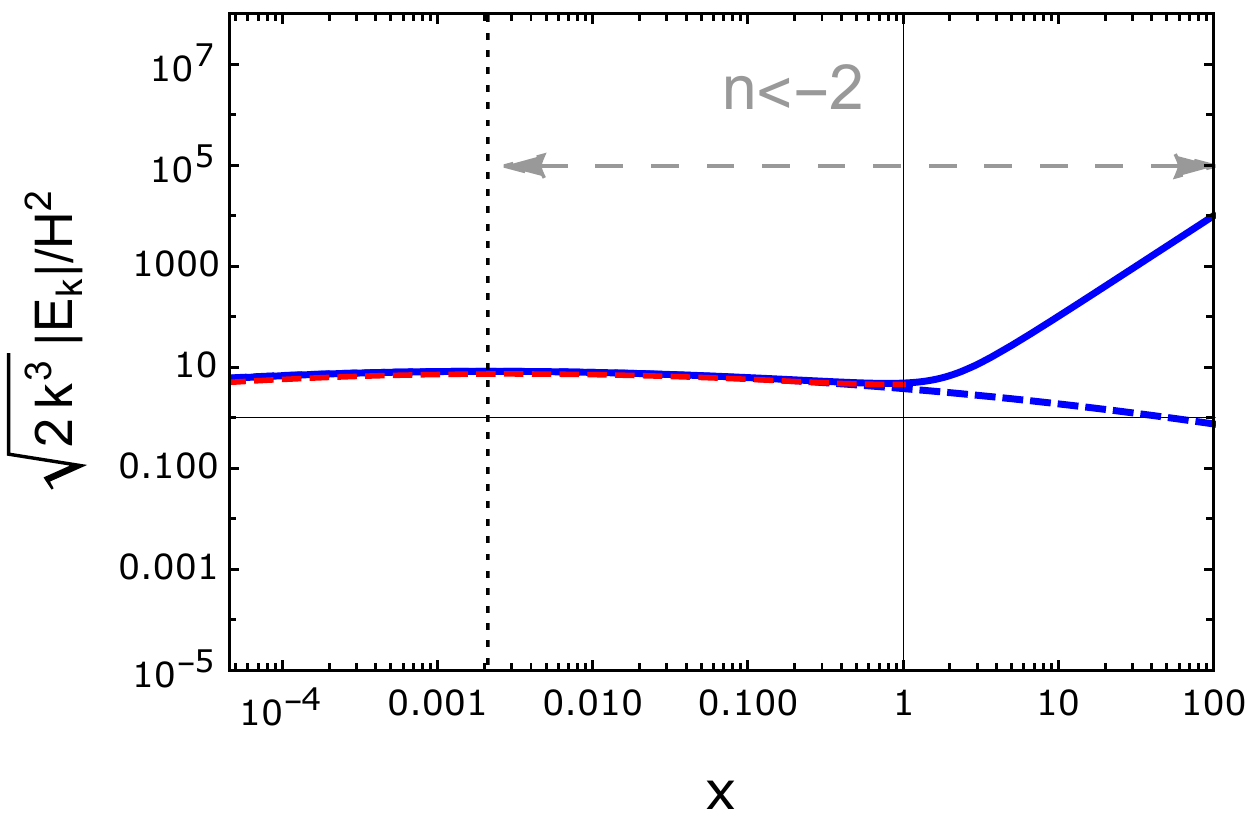}
  \end{minipage}
  \end{tabular}
  \caption{
  The time evolution of the electric field \eqref{eq:EkBk} with a mode exiting the horizon at $N = 10$ (left panel), $N=25$ (middle panel) and $N=50$ (right panel).
  The horizontal axis $x \equiv -k\tau$ is a dimensionless time flowing
  from the right to the left.
  The blue solid lines denote the exact numerical solutions.
  Regarding the initial conditions, we took Bunch-Davies vacuum and set the initial time variable as $x_{\rm ini} = 10^2$.
  The red dashed lines are the solutions with the uniform approximation \eqref{eq: Appro} which start to fit at around the horizon crossing $N = N_*$.
  The blue dashed lines are the approximate Gaussian fitting functions \eqref{eq: fit} discussed in section~\ref{tensormodes}.
  The black dotted lines denote a time when $n(t) = -2$.
  We have used the same parameter set as in Figure~\ref{fig:two}.
  }
 \label{fig:IA}
\end{figure}
%
%
\begin{align}
\langle \rho_{\gamma'} \rangle &= \langle\rho_{\gamma',T}\rangle + \langle\rho_{\gamma',L}\rangle \ , \\[1ex]
\langle\rho_{\gamma',T}\rangle &= \dfrac{1}{a^4}\int\dfrac{d\bm{k}}{(2\pi)^3}\left[ I^2\partial_\tau\left(\dfrac{V_k}{I}\right)\partial_\tau\left(\dfrac{V^*_k}{I}\right) + \left(k^2 + \dfrac{a^2m_{\gamma'}^2}{I^2}\right)V_kV_k^* \right] \nonumber \\
 &\equiv \int d\ln k \, \mathcal{P}_{\gamma',T}(k) \ , \\
\langle\rho_{\gamma',L}\rangle &= \dfrac{1}{2a^4}\int\dfrac{d\bm{k}}{(2\pi)^3}\left[ z_k^2\partial_\tau\left(\dfrac{X_k}{z_k}\right)\partial_\tau\left(\dfrac{X^*_k}{z_k}\right) + \left(k^2 + \dfrac{a^2m_{\gamma'}^2}{I^2}\right)X_kX_k^* \right] \nonumber \\
 &\equiv \int d\ln k \, \mathcal{P}_{\gamma',L}(k) \label{eq: lon} \ ,
\end{align}
where we have defined the power spectrum of the energy density
for each mode, i.e.~$P_{\gamma', T/L}$. Note that the two transverse polarization modes are summed over.
The gradient and mass terms are sub-dominant on the super-horizon scales in comparison with the kinetic terms, and we neglect their contributions to the energy density. 

Let us first evaluate the power spectrum for the longitudinal mode.
In our focused parameter space, $p \gg M$ always holds and therefore the solution of $X_k$ is given by Eq.~\eqref{eq: Xsol}.
In this case, the spectral shape is blue tilted and
only a few UV modes contribute to the energy density of the longitudinal mode
at the inflation end $N = N_{\rm end}$.
Its magnitude is evaluated as \cite{Nakai:2020cfw}
\begin{equation}
\langle\rho_{\gamma',L}\rangle|_{N_{\rm end}} \sim \dfrac{H^4}{8\pi^2} \, ,
\end{equation}
and found to be much smaller than the energy density of the transverse modes
as we will see below.

In considering the transverse modes on the occasion of negligible mass, it is convenient to define the corresponding (dark) electric and magnetic fields as $E^T_i = - I\partial_\tau A'^T_i/a^2$ and $B^T_i = I\epsilon_{ijk}\partial_j A'^T_k/a^2$, where $\epsilon_{ijk}$ is the Levi-Civita symbol in the flat spacetime.
Their mode functions can be defined as,
\footnote{ With $E_k$ and $B_k$, we can express $E_i^T$ and $B_i^T$ as
\begin{align}
    E_i^T & = \int \frac{d \bm{k}}{(2\pi)^3} \, e^{i \bm{k} \cdot \bm{x}} \left[ \left( E_k \, \hat{a}^X_{\bm{k}} + E_k^* \, \hat{a}^{X \dagger}_{-\bm{k}} \right) e_i^X( \hat{\bm{k}} ) + i \left( E_k \, \hat{a}^Y_{\bm{k}} + E_k^* \, \hat{a}^{Y \dagger}_{-\bm{k}} \right) e_i^Y( \hat{\bm{k}} ) \right] \; , \\
    B_i^T & = \int \frac{d \bm{k}}{(2\pi)^3} \, e^{i \bm{k} \cdot \bm{x}} \left[ \left( B_k \, \hat{a}^Y_{\bm{k}} + B_k^* \, \hat{a}^{Y \dagger}_{-\bm{k}} \right) e_i^X( \hat{\bm{k}} ) + i \left( B_k \, \hat{a}^X_{\bm{k}} + B_k^* \, \hat{a}^{X \dagger}_{-\bm{k}} \right) e_i^Y( \hat{\bm{k}} ) \right] \; ,
\end{align}
where we have used $\epsilon_{ijk} k_j e^X_k(\hat{\bm{k}}) = k e^Y_i(\hat{\bm{k}})$ and $\epsilon_{ijk} k_j e^Y_k(\hat{\bm{k}}) = - k e^X_i(\hat{\bm{k}})$. Be alert for the mixing of the polarizations $\{ X , Y \}$ between the operators and polarization vectors in the expression of $B_i^T$.
}
\begin{align}
E_k \equiv - \dfrac{I}{a^2}\dfrac{d}{d \tau}\left(\dfrac{V_k}{I}\right) \; , 
\qquad
B_k \equiv \frac{k}{a^2} \, V_k \; .
\label{eq:EkBk}
\end{align}
To evaluate these quantities, we use the result of the uniform approximation for $V_k$ obtained in \eqref{eq:V_UA}, or the corresponding super-horizon expression \eqref{eq:V_UA_approx}.
In the same way as in Appendix \ref{app:constantn}, the dark electric field $E_k$ is dominant over the magnetic counterpart $B_k$ in the case where the dark photon production occurs in the branch $n < 0$, and we neglect $B_k$ in comparison to $E_k$ in the following discussions.
In order to calculate $E_k$ under the uniform approximation, we define an ``averaged'' value of quantity $\nu(N)$ by
\begin{equation}
\bar{\nu}_k(N) \equiv \dfrac{\int_{N_*}^N d\tilde{N}\nu(\tilde{N})}{N-N_k} = -\left[ \dfrac{\tilde{\sigma}(N)-\tilde{\sigma}(N_*)}{c(N-N_k)} + \dfrac{1}{2}\dfrac{N-N_*}{N-N_k} \right] \ ,
\end{equation}
where $N_* \equiv N_k-\ln(\nu)$ is the number of e-foldings at which $g(\tau(N_*)) = 0$
and the definitions of $I(\sigma)$ and $n$, respectively \eqref{eq: I} and \eqref{eq: ndef}, have been used.
Then the following electric mode function obtained by the uniform approximation \eqref{eq:V_UA} and \eqref{eq:V_UA_approx},
\begin{align}
E^{\rm UA}_k &\equiv \dfrac{I}{a^2}\dfrac{d}{d \tau}\left(\dfrac{V^{\rm UA}_k}{I}\right) \notag \\
& \simeq e^{i\theta}\dfrac{3H^2}{\sqrt{2k^3}}\dfrac{2\sqrt{\nu}}{3}\exp\left[ \left(\bar{\nu}_k - \dfrac{3}{2}\right)(N-N_k) \right] \ ,
\quad 
(-k\tau \rightarrow 0) \, , 
\label{eq: Appro}
\end{align}
becomes a good approximation of the exact solution.
Figure~\ref{fig:IA} shows the time evolution of
the electric mode function for a few different momenta.
Recalling the e-folding times $N_1$ and $N_2$ defined below \eqref{eq:Vsigsig} such that $n(N_1) = n(N_2) = -2$ with $N_1 < N_2$, the left panel of Fig.~\ref{fig:IA}
shows an evolution of the electric field exiting the horizon before $N=N_1$.
For a while after leaving the horizon, its magnitude is suppressed until $|n|$ becomes greater than 
its critical value 2.
After $|n|$ exceeds 2, it starts to grow on the super-horizon regime.
Its growth persists until $N = N_2$ and after that the amplitude starts to decrease since $|n|$ becomes smaller than 2 again.
The middle panel gives an evolution of the mode function exiting the horizon just around $|n| = 2$.
The amplitude of the electric field with momentum modes of around this scale is mostly enhanced.
The right panel shows an evolution of the mode function exiting the horizon when $|n|$ is already greater than 2.
While it starts to grow, its magnitude is not maximally enhanced because the fluctuation is still in the sub-horizon for a while after $|n|$ crosses $2$.
We can see that the expression \eqref{eq: Appro} is well fitted to the numerical solution.

%
\begin{figure}[!t]
\center
  \includegraphics[width=90mm]{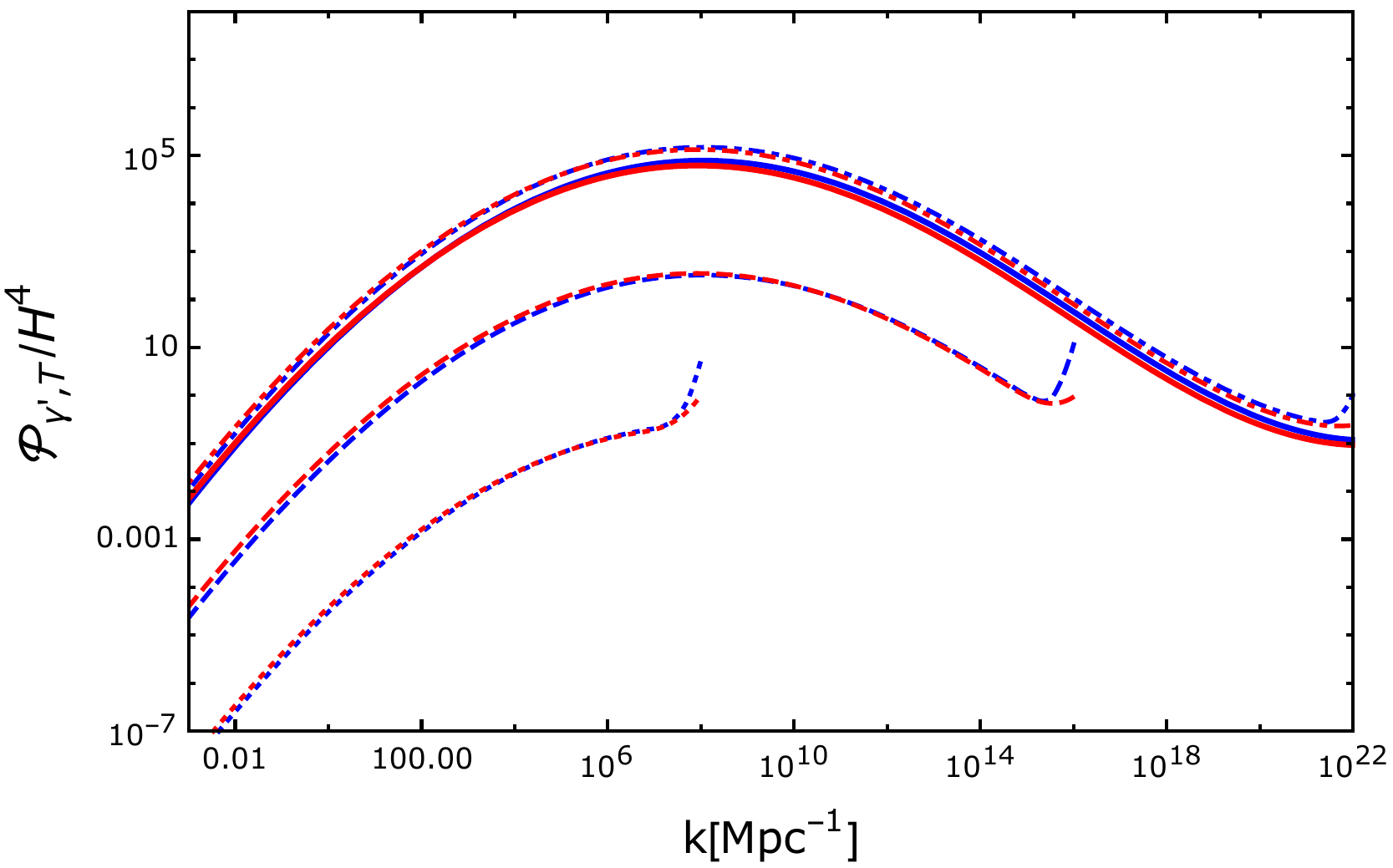}
  \caption{
  The power spectrum of the vector field $P_{\gamma',T}$ at e-folds $N = N_1$ (dotted), $N_*$ (dashed), $N_2$ (dot-dashed), $N_{\rm end}$ (solid).
  The blue and red lines denote the numerical and approximate solutions,
  respectively.
  The rapid increase for small scales represents the Bunch-Davies vacuum which should be renormalized as a UV contribution and we cut off its growth in our calculation.
  }
 \label{fig:PE}
\end{figure}
%

We now evaluate the time evolution of the power spectrum of the dark photon energy density with a certain model parameter set. The result is shown in Figure~\ref{fig:PE}.
The spectrum has a peak at an intermediate scale where the index $|n|$ becomes greater than 2.
This peak scale roughly 
corresponds to the number of e-foldings $N_k = N_1$, at which $|n|$ becomes equal to 2 for the first time during the evolution.
On the other hand, the amplitudes on large scales are suppressed because of a finite interval with $|n| < 2$ on the super-horizon regime (see Figure~\ref{fig:two}).
We will see in the next section that this feature makes it possible to avoid
the overproduction of isocurvature perturbation.
The present calculation assumes that the backreaction of the dark photon field
on the background motion of the spectator field is negligible.
As we will discuss in the next section, this assumption is justified
in our focused parameter region.

\section{Dark photon DM: constraints and results}
\label{DMabundance}

In this section, we estimate the relic abundance of the dark photon DM and
discuss theoretical and observational constraints on the current scenario.
Then, the viable parameter space of the dark photon DM is shown.

\subsection{Relic abundance}

Let us first derive the relic abundance of the dark photon DM,
\begin{equation}
\Omega_{\gamma'} = \dfrac{\langle\rho_{\gamma'}\rangle_{t=t_0}}{\rho(t_0)} \ ,
\end{equation}
where 
the average density of the dark photon is evaluated at the present time, denoted by subscript $0$, and
$\rho(t_0)$ is the critical density of our present universe: $\rho(t_0) = 3\Mpl^2H_0^2$.
To ensure the produced dark photon becomes non-relativistic some time after inflation ends at $t = t_{\rm end}$,
we compare the dark photon mass $m_{\gamma'}$ and
a physical, time-dependent momentum scale with a comoving wave number at which the spectrum is peaked, $q_{\rm peak}(t) \equiv k_{\rm peak}/a(t)$.
Using $k_{\rm peak} = a(t_{\rm peak})H(t_{\rm peak})$,
where the time $t_{\rm peak}$ is defined by this relation and is taken to be during inflation,
and defining a time duration of e-folds $\Delta N \equiv \log(a(t_{\rm end})/a(t_{\rm peak}))$,
the physical momentum scale $q_{\rm peak}(t)$ is written as
\begin{equation}
q_{\rm peak}(t) = H(t_{\rm peak}) e^{-\Delta N}\dfrac{a(t_{\rm end})}{a(t_{\rm reh})}\dfrac{a(t_{\rm reh})}{a(t)} \ ,
\end{equation}
where $t_{\rm reh}$ denotes the time when the reheating completes.
For simplicity, we assume an instantaneous reheating,
$a(t_{\rm end}) = a(t_{\rm reh})$, where the inflationary energy scale is related to the reheating temperature $T_{\rm reh}$ as
\begin{equation}
\rho(t_{\rm end}) = 3\Mpl^2H(t_{\rm end})^2 = \dfrac{\pi^2}{30}g_*(t_{\rm reh})T_{\rm reh}^4 \ . \label{eq: ins}
\end{equation}
Here, $g_*$ is the number of relativistic degrees of freedom.
The momentum scale of the dark photon at the reheating period is then evaluated as
\begin{equation}
q_{\rm peak}(t_{\rm reh}) \sim 
0.1 \, \text{GeV}
\left(\dfrac{H_{\rm inf}}{10^{12} \ \text{GeV}}\right)\left(\dfrac{e^{30}}{e^{\Delta N}}\right) \ ,
\end{equation}
where $H_{\rm inf}$ denotes the inflationary Hubble scale.
In our scenario, we assume a high-scale,
almost de-Sitter inflation and
$H(t_{\rm peak}) \simeq H(t_{\rm end}) \simeq H_{\rm inf}$.
Therefore, the dark photon with mass $m_{\gamma'} \ll \text{eV}$ in our interest
is relativistic right after inflation ends.
Then, the current energy density of the dark photon evolves from the end of inflation as
\begin{equation}
\langle\rho_{\gamma'}\rangle_{t=t_0} = \langle\rho_{\gamma'}\rangle_{t=t_{\rm end}}\left(\dfrac{a_{\rm reh}}{a(t_{\rm NR})}\right)^4\left(\dfrac{a(t_{\rm NR})}{a_0}\right)^3 \ ,
\end{equation}
where $t_{\rm NR}$ is the time when the dark photon becomes non-relativistic, determined by
\begin{equation}
q_{\rm peak}(t_{\rm NR}) = m_{\gamma'} \ .
\end{equation}
By using the entropy conservation law, the present abundance of the dark photon DM
is obtained as
\begin{equation}
\Omega_{\gamma'} = \left(\dfrac{\pi^2}{90}\right)^{3/2}\dfrac{g_S(t_0)g_*(t_{\rm reh})^{3/2}}{g_S(t_{\rm reh})}\dfrac{T_{\rm reh}^3}{\Mpl^3}\dfrac{m_{\gamma'}T_0^3}{3\Mpl^2H_0^2}\dfrac{\langle\rho_{\gamma'}\rangle_{t=t_{\rm end}}}{H_{\rm inf}^4}\dfrac{1}{-k_{\rm peak}\tau_{\rm end}} \ .
\end{equation}
Therefore, the mass of the dark photon is evaluated as
\begin{equation}
m_{\gamma'} \simeq 
8.2 \times 10^{-11} \, {\rm eV}
\left( \frac{\Omega_{\gamma'} h^2}{0.14} \right)
\left(\dfrac{10^{7}}{\langle\rho_{\gamma'}\rangle_{t=t_{\rm end}}/H_{\rm inf}^4}\right)\left(\dfrac{e^{30}}{e^{\Delta N}}\right)\left(\dfrac{10^{15}\text{GeV}}{T_{\rm reh}}\right)^{3} \ , \label{eq: mass}
\end{equation}
where we have used $H_0 = 2.133\times10^{-42} \, h \, \text{GeV}$, $T_0 = 2.725 \, \text{K} = 2.348\times10^{-13} \,\text{GeV}$, $\Mpl = 2.435\times10^{18}\text{GeV}$
and $g_S(t_0) = 3.91$,
and assumed $g_*(t_{\rm reh}) = g_S(t_{\rm reh}) = 106.75$.
From this expression, we can observe that, in order for the dark photon that explains the total dark matter abundance to have a smaller mass, either its energy density $\rho_{\gamma'}$ during inflation , $\Delta N$ or the reheating temperature $T_{\rm reh}$ must be larger. The choice of these parameters in \eqref{eq: mass} is rather an optimistic one, in favor of small mass, and thus it gives a rough lower bound for the dark photon mass $m_{\gamma'}$ in our model. Yet we remark that it is sensitive to the values of $\Delta N$ and $T_{\rm reh}$, and a mass smaller than $\sim 10^{-10} \, {\rm eV}$ by a few orders of magnitude is still feasible.

\subsection{Constraints}

Next, we discuss relevant theoretical
and observational constraints on the present scenario.

\subsubsection*{(i) Isocurvature mode}

Let us consider the isocurvature mode of the dark photon in our scenario, as it could be one of the stringent constraints on the inflationary production of dark photon DM \cite{Nakayama:2020rka}.
The isocurvature fluctuation of the dark photon is defined as the entropy perturbation caused by the non-adiabatic mode:
\begin{equation}
\mathcal{S} \equiv \dfrac{\delta\rho_{\gamma'}}{\langle \rho_{\gamma'} \rangle} = \dfrac{\rho_{\gamma'} - \langle \rho_{\gamma'} \rangle}{\langle \rho_{\gamma'} \rangle} \ .
\end{equation}
Then, in terms of the Fourier decomposition,
\begin{equation}
\frac{\rho_{\gamma'}(\bm{x})}{\langle \rho_{\gamma'} \rangle}
\equiv \int \frac{d \bm{k}}{(2\pi)^3} \, \hat\delta_{\gamma'}(\bm{k}) \, e^{i \bm{k} \cdot \bm{x}}
\end{equation}
its power spectrum is given by
\begin{align}
\langle \mathcal{S}(\bm{x})\mathcal{S}(\bm{y}) \rangle &= \int\dfrac{d\bm{k}d\bm{k}'}{(2\pi)^6}
\left[ \langle \hat\delta_{\gamma'} (\bm{k}) \, \hat\delta_{\gamma'}(\bm{k}') \rangle - 
\left( 2 \pi \right)^6 \delta (\bm{k}) \, \delta(\bm{k}')
\right]
e^{i\bm{k}\cdot\bm{x} + \bm{k}'\cdot\bm{y}} \\
&= \int \dfrac{dk}{k} \dfrac{\sin(k|\bm{x}-\bm{y}|)}{k|\bm{x}-\bm{y}|} \, \mathcal{P}_{\mathcal{S}}(k) \ ,
\end{align}
where the power spectrum $\mathcal{P}_S$ is defined through
\begin{equation}
\langle \hat\delta_{\gamma'} (\bm{k}) \, \hat\delta_{\gamma'}(\bm{k}') \rangle - \left( 2 \pi \right)^6 \delta(\bm{k}) \, \delta(\bm{k}')
= (2\pi)^3\delta(\bm{k} + \bm{k}')\dfrac{2\pi^2}{k^3}\mathcal{P}_\mathcal{S}(k) \ .
\label{eq:def_PS}
\end{equation}
Note that the disconnected contribution is explicitly subtracted in the expression \eqref{eq:def_PS}.
Since the transverse electric mode is energy-dominant, $\hat\delta_{\gamma'}(\bm{k})$ is approximately given by
\begin{equation}
\hat\delta_{\gamma'}(\bm{k}) 
\simeq \dfrac{1}{2\langle \rho_{\gamma',T} \rangle}\int\dfrac{d\bm p}{(2\pi)^3}\left(\hat{E}^X_{\bm{p}}e^X_{i}(\hat{\bm{p}}) + i\hat{E}^Y_{\bm{p}}e^Y_{i}(\hat{\bm{p}}) \right)\left(\hat{E}^X_{\bm{k}-\bm{p}}e^X_i(\widehat{\bm{k}-\bm{p}}) + i\hat{E}^Y_{\bm{k}-\bm{p}}e^Y_i(\widehat{\bm{k}-\bm{p}}) \right) \ .
\end{equation}
A hat on the electric field manifests that 
$\hat{E}^X_{\bm{p}} \equiv E_p \, \hat{a}^X_{\bm{p}} + E_p^* \, \hat{a}^{X \dagger}_{-\bm{p}}$
is an operator. 
Then, noting $e^X_i(\hat{\bm{p}})e^Y_i(\widehat{\bm{k}-\bm{p}}) = e^Y_i(\hat{\bm{p}})e^X_i(\widehat{\bm{k}-\bm{p}}) = 0$,%
\footnote{This is because the vectors $e^{X}_i(\hat{\bm{p}})$ and $e^{X}_i (\widehat{\bm{k}-\bm{p}})$ are both on the plane defined by $\bm{k}$ and $\bm{p}$, and $e^{Y}_i(\hat{\bm{p}})$ and $e^{Y}_i(\widehat{\bm{k}-\bm{p}})$ are both perpendicular to it, under our construction of the polarization vectors.}
the power spectrum is evaluated as
\begin{equation}
\mathcal{P}_{\mathcal{S}}(k) \simeq \dfrac{1}{4}\sum_{\sigma=X,Y}\dfrac{1}{\pi^2\langle\rho_{\gamma',T}\rangle^2}\int\dfrac{d\bm{p}_*}{(2\pi)^3}\left|e^\sigma_i(\hat{\bm{p}})e^\sigma_i(\widehat{\bm{k}-\bm{p}})\right|^2|E_{|\bm{p}|_*}|^2|E_{|\bm{k}-\bm{p}|_*}|^2 \ , \label{eq: iso}
\end{equation}
where we have introduced the dimensionless momenta: $\bm{p}_* \equiv \bm{p}/k, \ |\bm{k}-\bm{p}|_* \equiv |\bm{k}-\bm{p}|/k$.
Note that the disconnected term in $\langle \hat\delta_{\gamma'} (\bm{k}) \hat\delta_{\gamma'}(\bm{k}') \rangle$, which is proportional to $\delta(\bm{k}) \delta(\bm{k}')$, exactly cancels with the corresponding term in \eqref{eq:def_PS}, and thus $\mathcal{P}_S$ takes into account only the connected diagrams.
The amount of the isocurvature perturbation is tightly constrained at CMB scales.
The current {\it Planck} satellite observation puts a constraint on the magnitude of the isocurvature power spectrum evaluated at three different scales $k = 0.002, \ 0.05, \ 0.1 \ \text{Mpc}^{-1}$.
As shown in Figure~\ref{fig:PE}, the resultant spectral shape is blue-tilted
at CMB scales in our model.
To constrain the amplitude of the power spectrum, we make use of the limits evaluated at the lowest and highest scales,
$k = 0.002 \ \text{Mpc}^{-1}$ and $k = 0.1 \ \text{Mpc}^{-1}$, which are given by \cite{Planck:2018jri}
\begin{equation}
\mathcal{P}_{\mathcal{S}}|_{k=0.002\text{Mpc}^{-1}} \lesssim 0.8\times10^{-10} \ , \qquad \mathcal{P}_{\mathcal{S}}|_{k=0.1\text{Mpc}^{-1}} \lesssim 2.2\times10^{-9} \ . \label{eq: isocon}
\end{equation}
The spectrum in our model has a peaky feature, and thus imposing the above observational isocurvature constraints on the peak value should provide a conservative upper bound on the dark photon production.

\subsubsection*{(ii) Backreaction}

The Friedmann equation,
\begin{equation}
3\Mpl^2H^2 = \dfrac{1}{2}\dot{\phi}^2 + U(\phi) + \dfrac{1}{2}\dot{\sigma}^2 + V(\sigma) + \langle \rho_{\gamma'} \rangle \, ,
\end{equation}
where $U(\phi)$ denotes the inflaton potential,
includes the contribution from the energy density of the produced dark photon.
Our calculations in this paper relies on the assumption that the inflationary quasi-de-Sitter background is driven only by $\phi$, and the produced dark photon has negligible impact on the dynamics of $\sigma$, which is already subdominant to $\phi$.
To neglect the backreaction contribution, we need
\begin{equation}
\langle \rho_{\gamma'} \rangle \ll \frac{1}{2} \, \dot\sigma^2 \ll 3 \Mpl^2 H^2 \; ,
\end{equation}
during inflation.
This constraint is not tighter than that from the backreaction to the motion of the spectator field in Eq.~\eqref{eq: sigmaeom}.
Namely, it suffices that the following condition must be satisfied:
\begin{equation}
\left|3\dfrac{d\sigma}{dN}\right| \gg \left|-\dfrac{II_\sigma}{2H^2}\langle F'_{\mu\nu}F'^{\mu\nu} \rangle \right| \simeq \dfrac{2\langle\rho_{\gamma',T}\rangle}{H^2\Lambda_1} \, ,
\end{equation}
where in the most right-hand side we have ignored the contribution of the longitudinal mode.
To characterize this hierarchy, we define a ratio,
\begin{equation}
R_b \equiv \left|\dfrac{2\langle\rho_{\gamma',T}\rangle}{3H\Lambda_1\dot{\sigma}}\right| \, ,
\end{equation}
and consider the parameter region realizing $R_b \ll 1$.
By using $\dot{\sigma}/(H\Lambda_1) \simeq n$ in Eq.~\eqref{eq: na}
and the condition of slow-roll parameters \eqref{eq: slow1} with \eqref{eq: slow2},
\begin{equation}
\epsilon_\sigma < \epsilon_H \qquad \longleftrightarrow \qquad \dfrac{\Lambda_1^2}{2\Mpl^2} < \dfrac{\epsilon_H}{n^2} \, ,
\end{equation}
on CMB scales, we can translate the condition $R_b \ll 1$ to
the following constraint on
the magnitude of the dark photon energy density: 
\begin{equation}
\frac{\langle \rho_{\gamma', T} \rangle}{H^4} \ll \frac{3\vert n \vert}{2} \, \frac{\Lambda_1^2}{H^2} < \dfrac{3|n|}{8\pi^2n_{\rm CMB}^2\mathcal{P}_{\zeta, \rm CMB}} \sim 2\times10^7|n| \ ,
\label{eq:rhogamma_upperbound}
\end{equation}
where the power spectrum of the curvature perturbation in the vacuum state is given by
\begin{equation}
\mathcal{P}_{\zeta, \rm CMB} = \dfrac{H^2}{8\pi^2\Mpl^2\epsilon_H} \simeq 2\times10^{-9} \ .
\end{equation}
Note that in \eqref{eq:rhogamma_upperbound} we distinguish $n$ from $n_{\rm CMB}$, the latter denoting the value of $n$ at the time when the CMB modes exit the horizon. In our mechanism, the value of $n$ changes during inflation by an $\mathcal{O}(1)$ amount, and specifically we demand $\vert n_{\rm CMB} \vert < 2$ in order to avoid overproduction of dark photon spoiling the CMB predictions. Typical values of $n_{\rm CMB}$ in our considerations are $\sim 1$, which is taken for the final evaluation in \eqref{eq:rhogamma_upperbound}.

\subsubsection*{(iii) Non-relativistic time vs.~equality time}

After inflation ends, the dark photon behaves as a relativistic radiation component since the scale of the physical momentum at the spectral peak $q_{\rm peak}$ is much larger than the dark photon mass $m_{\gamma'}$.
For the dark photon to behave as a viable dark matter candidate, it must become non-relativistic 
by the time of matter-radiation equality $t = t_{\rm eq}$.
We define the cosmic temperature $T_{\rm NR}$ at time $t=t_{\rm NR}$ when the scale of the physical momentum equals to the dark photon mass,
\begin{equation}
q_{\rm peak}(t_{\rm NR}) = \dfrac{k_{\rm peak}}{a(t_{\rm NR})} = m_{\gamma'} \ .
\end{equation}
Using the entropy-conservation law, $g_Sa^3T^3 = \text{const}.$, the temperature $T_{\rm NR}$ is obtained as\footnote{Ref.~\cite{Nakai:2020cfw} has missed the factor $(g_S(t_{\rm reh})/g_S(t_{\rm NR}))^{1/3}$, while the effect is small.}
\begin{align}
T_{\rm NR} &= m_{\gamma'} \, \dfrac{T_{\rm reh}}{q_{\rm peak}(t_{\rm reh})}\left(\dfrac{g_S(t_{\rm reh})}{g_S(t_{\rm NR})}\right)^{1/3} \notag \\
&= \left(\dfrac{30}{\pi^2g_*(t_{\rm reh})}\right)^{1/4}\dfrac{m_{\gamma'}}{H(t_{\rm end})}\left(\dfrac{\rho(t_{\rm end})}{\rho(t_{\rm reh})}\right)^{1/12}\dfrac{\rho(t_{\rm end})^{1/4}}{-k_{\rm peak}\tau_{\rm end}}\left(\dfrac{g_S(t_{\rm reh})}{g_S(t_{\rm NR})}\right)^{1/3} \ ,
\end{align}
assuming that the equation of state is of matter domination after inflation and before reheating.
On the other hand, the temperature $T_{\rm eq}$ at the equality time is determined by solving the energy density equality between the radiation and matter components,
\begin{equation}
\dfrac{\Omega_r}{a(t_{\rm eq})^4} = \dfrac{\Omega_m}{a(t_{\rm eq})^3} \ ,
\end{equation}
with the normalization of the scale factor now taken at present, $a(t_0) = 1$.
Then, the entropy-conservation law and  $\Omega_r = \pi^2g_*(t_0)T_0^4/(90\Mpl^2H_0^2)$ give
\begin{equation}
T_{\rm eq} = \dfrac{90\Omega_m\Mpl^2H_0^2}{\pi^2g_*(t_0)T_0^3} \ .
\end{equation}
Imposing $T_{\rm NR} > T_{\rm eq}$, we obtain the following constraint:
\begin{equation}
\dfrac{T_{\rm NR}}{T_{\rm eq}} = \dfrac{(\pi^2/90)^{1/2}}{\Omega_m} \, \dfrac{g_*(t_0)}{g_*(t_{\rm reh})^{1/2}} \, \dfrac{m_{\gamma'}}{T_{\rm reh}} \, \dfrac{T_0^3}{\Mpl H_0^2} \, e^{\Delta N}\left(\dfrac{g_S(t_{\rm reh})}{g_S(t_{\rm NR})}\right)^{1/3} > 1 \ , \label{eq: con1}
\end{equation}
where the instantaneous reheating is assumed. Henceforth we take $g_S(t_{\rm eq}) = g_S(t_0) = 3.909, \ g_*(t_{\rm eq}) = g_*(t_0) = 3.363$.
Defining $\Delta N = \Delta N_{\rm eq}$ at which $T_{\rm NR} = T_{\rm eq}$, Eq.~\eqref{eq: con1} is rewritten as
\begin{equation}
\Delta N > \Delta N_{\rm eq} \simeq 10.5 
+ \ln\left(\dfrac{\Omega_m h^2}{0.143}\right)
+ \ln\left(\dfrac{T_{\rm reh}}{10^{13} \ \text{GeV}}\right) - \ln\left(\dfrac{m_{\gamma'}}{10^{-10} \ \text{eV} }\right) \ . \label{eq: neqcons}
\end{equation}
We have by now collected all the constraints that are to be imposed on the amount of the production of dark photon dark matter in our model.

\subsection{Results}
\label{result}

\begin{figure}[tbp]
\begin{minipage}{0.5\hsize}
\begin{center}
  \includegraphics[width=70mm]{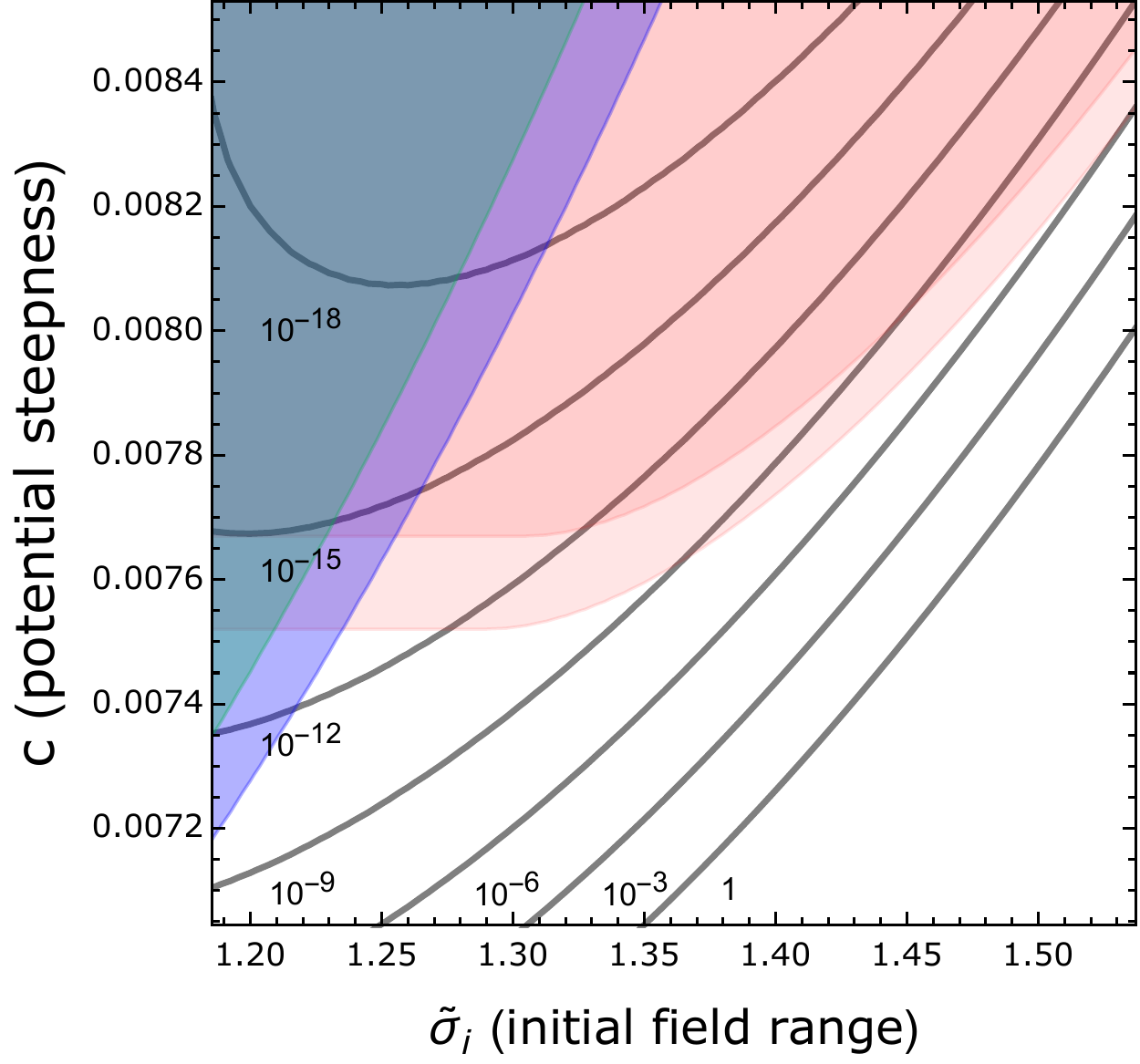}
  \end{center}
  \end{minipage}
  \begin{minipage}{0.5\hsize}
\begin{center}
  \includegraphics[width=70mm]{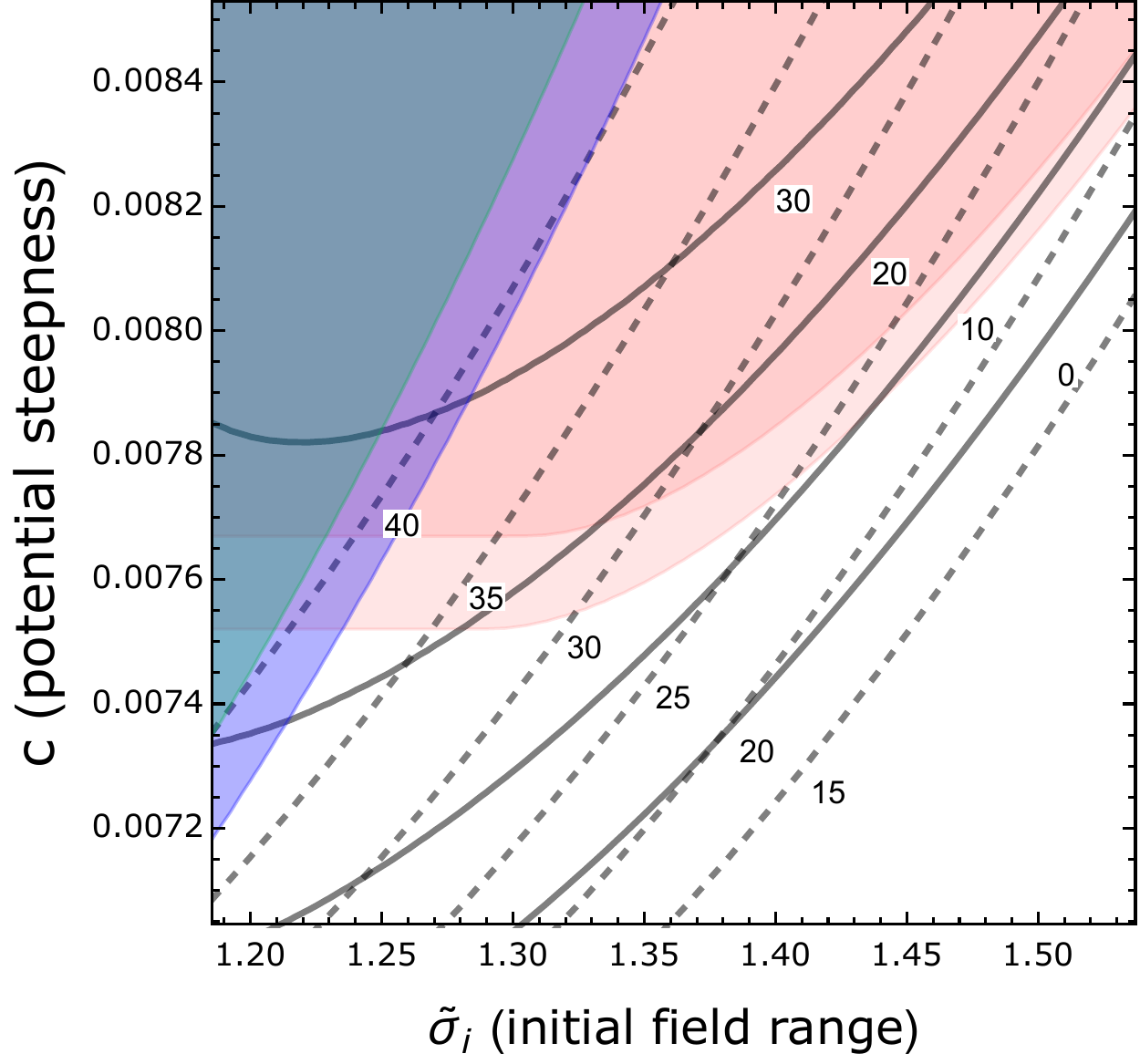}
  \end{center}
  \end{minipage}
  \caption{
  {\it Left panel} : contour plots of the dark photon DM with mass $m_{\gamma'} = 1$, $10^{-3}$, $10^{-6}$, $\ 10^{-9}$, $10^{-12}$, $10^{-15}$, $10^{-18} \ \text{eV}$ (black lines)
  in the parameter space of $\tilde{\sigma}_i$ and $c$.
  {\it Right panel} : contour plots of $\Delta N_{\rm eq} = 0, 10, 20, 30$ (black solid lines) and $\Delta N = 15, 20, 25, 30, 35, 40$ (black dashed lines). In both panels, we set $R = 10^{-2}$, $\Lambda_1 = 5\times10^{-3}M_{\rm Pl}$ and $r_{ \rm v} = 5\times10^{-4}$.
  The shaded regions are constrained by the backreaction,
  $R_b>0.1$ (light red) and $R_b > 1$ (red), and
  the isocurvature perturbation with $k = 0.002 \ \text{Mpc}^{-1}$ (dark green) and $k=0.1 \ \text{Mpc}^{-1}$ (purple).
  }
 \label{fig:result}
\end{figure}

We numerically find a viable parameter region of the dark photon DM evaluated in \eqref{eq: mass} consistent with the constraints discussed above, which is shown in Figure~\ref{fig:result}.
To make this plot, we have solved the (massless approximate) equation of motion for dark photon, \eqref{eq: Ve3}, 
together with the equation of motion for $\sigma$, \eqref{eq: sigmaeom}, with constant $H$,
with several values of $\tilde{\sigma}_i$ and $c$, and used the resultant solution of $\rho_{\gamma',T}$ shown in Figure \ref{fig:PE}.
The initial field range $\tilde{\sigma}_i$ characterizes a timing when $|n|$ becomes greater than $2$.
For a small $\tilde{\sigma}_i$, $|n|$ crosses $2$ at an early stage of inflation, which leads to a large $\Delta N$ and correspondingly a small $m_{\gamma'}$
for the dark photon DM.
However, the effect of the isocurvature mode is severe for this case because the dark photon is amplified on large scales.
For a large $\tilde{\sigma}_i$, $|n|$ crosses $2$ at a late stage of inflation, which leads to a large $m_{\gamma'}$.
The effect of backreaction tends to reduce in this case 
because $N_2$ becomes closer to $N_{\rm end}$ or even exceeds it, and therefore the time interval of particle production becomes shorter.
On the other hand, the parameter $c$ is related to the steepness of the potential slope.
As $c$ increases, the slope of the potential becomes steeper and
the value of $|n|$ gets larger.
As a result, the effect of the backreaction and/or the contribution of the isocurvature mode become severer.
However, a large $c$ and a small $\tilde{\sigma}_i$ tend to derive the dynamics of $|n|$ damping so early and therefore predict a small $\langle \rho_{\gamma',T} \rangle$ leading to a large $m_{\gamma'}$.
To find a viable region for the light dark photon DM, a high-scale inflation would be preferable because a high reheating temperature is required
in Eq.~\eqref{eq: mass}.
We can see from the figure that the present scenario predicts a mass window $m_{\gamma'} \gtrsim 10^{-13} \ \text{eV}$.
Due to the exponential sensitivity to the variation of model parameters,
the viable parameter space leading to a preferable range of mass 
$m_{\gamma'}$ for light dark photon is localized in a small region
on the $\tilde\sigma_i \, \mbox{-} \, c$ plane.
The figure also describes the constraint that the dark photon must be non-relativistic until the equality time.
We find that in the region of our interest the condition \eqref{eq: neqcons}
is satisfied.

In our study, we have assumed that the Hubble parameter $H$ is constant in the whole period of inflation and therefore neglected its dynamics near and after the end of inflation.
However, when we take it into account,
we might expect that the allowed region of the dark photon mass
would get wider, and even a smaller mass would be available.
This is because the amplitude of the energy density of the dark photon is
enhanced due to the rapid motion of the spectator field.
However, the backreaction effect 
may become dominant and the system can be completely non-linear.
Such a near-end and post-inflationary amplification is expected to be severe if the potential of $\sigma$ becomes flat after inflation, unlike our choice of $V(\sigma)$ as in \eqref{eq: tanh2}.
The analysis of such a non-linear system is beyond the scope of the present work.

%
%

\section{Generation of tensor modes}\label{tensormodes}

In this section, we evaluate the power spectrum of tensor modes sourced by the dark photon field during inflation.
This effect is an inevitable consequence of the dark photon production and is potentially led to observable signals, which we would like to evaluate in this section.
The tensor perturbation is given by fluctuations of the spacial components of the metric, $g_{ij}(t, \bm{x}) = a(t)^2(\delta_{ij} + \tfrac{1}{2}h_{ij}(t,\bm{x}))$, which obey the following equation of motion at the leading order:
\begin{equation}
\left[ \partial_t^2+3H\partial_t - \dfrac{\nabla^2}{a^2} \right]h_{ij} \simeq -\dfrac{4}{\Mpl^2}\Pi^{lm}_{ij}E_lE_m \ ,
\label{eq:eom_tensor}
\end{equation}
where $\Pi^{lm}_{ij}$ is the transverse-traceless projector defined by
\begin{equation}
\Pi^{lm}_{ij} \equiv \Pi^l_i\Pi^m_j - \dfrac{1}{2}\Pi_{ij}\Pi^{lm} \ , \qquad \Pi_{ij} \equiv \delta_{ij} - \dfrac{\partial_i\partial_j}{\nabla^2} \ .
\end{equation}
We decompose $h_{ij}$ into the linear polarization tensors in Fourier space,
\begin{align}
h_{ij}(t,\bm{x}) &= \int\dfrac{d\bm{k}}{(2\pi)^3}\hat{h}_{ij}(\bm{k}, t)e^{i\bm{k}\cdot\bm{x}} 
= \sum_{s = +,\times}\int\dfrac{d\bm{k}}{(2\pi)^3}e^s_{ij}(\hat{\bm{k}})\hat{h}^s_{\bm
{k}}(t)e^{i\bm{k}\cdot\bm{x}} \ ,
\end{align}
where the transverse-traceless polarization tensors $e^{+/\times}_{ij}(\hat{\bm{k}})$ are given by the following products of the polarization vectors $e^{X/Y}_i(\hat{\bm{k}})$:
\begin{align}
e^+_{ij}(\hat{\bm k}) &= \dfrac{1}{\sqrt{2}}\left( e^X_i(\hat{\bm k})e^X_j(\hat{\bm k}) - e^Y_i(\hat{\bm k})e^Y_j(\hat{\bm k}) \right) \ , \\
e^\times_{ij}(\hat{\bm k}) &= \dfrac{i}{\sqrt{2}} \left( e^X_i(\hat{\bm k})e^Y_j(\hat{\bm k}) + e^Y_i(\hat{\bm k})e^X_j(\hat{\bm k}) \right) \ ,
\end{align}
so that they satisfy the property $e_{ij}^{s \, *}(-\hat{\bm{k}}) = e_{ij}^s(\hat{\bm{k}}).$
Then, by using $\hat{h}^s_{\bm
{k}} = e^{s \, *}_{ij}(\hat{\bm k}) \, \hat{h}_{ij}(\hat{\bm k})$
and $\Pi_{ij}^{lm}e^s_{lm}(\hat{\bm k}) = e^s_{ij}(\hat{\bm k})$,
we obtain
\begin{align}
\left[ \partial_x^2 +1 - \dfrac{2}{x^2} \right](a\hat h^{s}_{\bm k}) = &
-
e^{s \, *}_{ij}(\hat{\bm{k}}) \,
\dfrac{4a^3}{k^2\Mpl^2} \notag \\
&\times \int\dfrac{d\bm p}{(2\pi)^3}\left(\hat{E}^X_{\bm{p}}e^X_{i}(\hat{\bm{p}}) + i\hat{E}^Y_{\bm{p}}e^Y_{i}(\hat{\bm{p}}) \right) \nonumber \\
&\qquad \qquad \times \left(\hat{E}^X_{\bm{k}-\bm{p}}e^X_{j}(\widehat{\bm{k}-\bm{p}}) + i\hat{E}^Y_{\bm{k}-\bm{p}}e^Y_{j}(\widehat{\bm{k}-\bm{p}}) \right) \ . \label{eq: ten}
\end{align}
on the de Sitter assumption.
The solution to this equation is a linear combination of two components,
$\hat{h}^s_{\bm k} = \hat{h}^s_{\bm k, \rm v} + \hat{h}^s_{\bm k, \rm s}$.
The homogeneous solution $\hat{h}^s_{\bm k, \rm v}$ is the usual Banch-Davies vacuum mode and its mode function is given by,
at the leading order in slow roll,
\begin{equation}
a h^s_{k, \rm v} = \dfrac{e^{-ik\tau}}{
\Mpl
\sqrt{2k}}\left(1 - \dfrac{i}{k\tau} \right) \ ,
\end{equation}
and the corresponding tensor-to-scalar ratio reads
\begin{align}
r_{v} \equiv \frac{8 \mathcal{P}_{h,{\rm v}}}{\mathcal{P}_{\zeta,\rm CMB}} = 16 \, \epsilon_H \; .
\end{align}
The numerical factor $8$ in the intermediate step comes from our definition $h_{ij} = 2 \delta g_{ij} /a^2$ and from the $2$ polarization states of the tensor modes.
On the other hand, the peculiar solution $\hat{h}^s_{\bm k, \rm s}$ is sourced by the second order
of the dark photon field.
We find the mode function by using the Green function's method,
\begin{align}
a\hat{h}^s_{\bm{k},s} &= -\dfrac{4}{k^2\Mpl^2} \,
e^{s \, *}_{ij}(\hat{\bm{k}})
\int_{-\infty}^{\infty} d y ~a^3(y) \, G_R(x, \ y) \notag \\
&\quad \times\int\dfrac{d\bm p}{(2\pi)^3}\left(\hat{E}^X_{\bm{p}}(y) \, e^X_{i}(\hat{\bm{p}}) 
+ i\hat{E}^Y_{\bm{p}}(y) \, e^Y_{i}(\hat{\bm{p}}) \right)
\left(\hat{E}^X_{\bm{k}-\bm{p}}(y) \, e^X_{j}(\widehat{\bm{k}-\bm{p}}) 
+ i\hat{E}^Y_{\bm{k}-\bm{p}}(y) \, e^Y_{j}(\widehat{\bm{k}-\bm{p}}) \right) \ ,
\end{align}
where $G_R(x,y) \equiv
\Theta(y-x) \left[ (x-y) \cos(x-y) - (1+xy) \sin(x-y) \right]/(x y) \simeq
-\Theta(y-x)(x^3-y^3)/(3xy)$ is the retarded Green function,
where the last approximate equality is in the limit of $x \equiv - k \tau \ll 1$ and $y \equiv - k \tau' \ll 1$, with $\tau'$ an auxiliary time variable.
This approximation is valid thanks to the fact that our mechanism of dark photon amplification takes place on the super-horizon scales.
Regarding the products of the polarization tensors and vectors,
we assume $\bm k$ is directed in the $\hat{z}$ axis
and use the following identities:
\begin{eqnarray}
&&e^{+*}_{ij}(\hat{\bm{k}})e^{X}_{i}(\hat{\bm{p}})e^{X}_{j}(\widehat{\bm{k}-\bm{p}}) = -\dfrac{\cos\theta_{\hat{\bm p}}\cos\theta_{\widehat{\bm k - \bm p}}\cos 2\phi_{\hat{\bm p}}}{\sqrt 2} \, , \\
&&e^{+*}_{ij}(\hat{\bm{k}})e^{Y}_{i}(\hat{\bm{p}})e^{Y}_{j}(\widehat{\bm{k}-\bm{p}}) = \dfrac{\cos 2\phi_{\hat{\bm p}}}{\sqrt 2} \label{eq: id4} \ , \\
&&e^{+*}_{ij}(\hat{\bm{k}})e^{X}_{i}(\hat{\bm{p}})e^{Y}_{j}(\widehat{\bm{k}-\bm{p}}) = 
\frac{\cos \theta_{\hat{\bm{p}}} \sin 2 \phi_{\hat{\bm{p}}}}{\sqrt{2}}
\, , \\
&&e^{+*}_{ij}(\hat{\bm{k}})e^{Y}_{i}(\hat{\bm{p}})e^{X}_{j}(\widehat{\bm{k}-\bm{p}}) = 
\frac{\cos\theta_{\widehat{\bm k - \bm p}} \sin 2 \phi_{\hat{\bm{p}}}}{\sqrt{2}}
\label{eq: id5} \ , \\
&&e^{\times*}_{ij}(\hat{\bm{k}})e^{X}_{i}(\hat{\bm{p}})e^{X}_{j}(\widehat{\bm{k}-\bm{p}}) = 
i\dfrac{
\cos\theta_{\widehat{\bm p}}\cos\theta_{\widehat{\bm k - \bm p}} \sin 2 \phi_{\hat{\bm p}}}{\sqrt 2} \, , \\
&&e^{\times*}_{ij}(\hat{\bm{k}})e^{Y}_{i}(\hat{\bm{p}})e^{Y}_{j}(\widehat{\bm{k}-\bm{p}}) = 
\dfrac{
\sin 2 \phi_{\hat{\bm p}}}{\sqrt{2} i}
\label{eq: id7} \ , \\
&&e^{\times*}_{ij}(\hat{\bm{k}})e^{X}_{i}(\hat{\bm{p}})e^{Y}_{j}(\widehat{\bm{k}-\bm{p}}) = i\dfrac{
\cos\theta_{\hat{\bm p}} 
\cos 2 \phi_{\hat{\bm p}}}{\sqrt 2} \, , \\
&&e^{\times*}_{ij}(\hat{\bm{k}})e^{Y}_{i}(\hat{\bm{p}})e^{X}_{j}(\widehat{\bm{k}-\bm{p}}) = i\dfrac{
\cos\theta_{\widehat{\bm k - \bm p}} 
\cos 2 \phi_{\hat{\bm p}}}{\sqrt 2}
\label{eq: id6} \ ,
\end{eqnarray}
with $\cos\theta_{\hat{\bm p}} = \hat{\bm{k}} \cdot \hat{\bm{p}}$ and $\cos\theta_{\widehat{\bm k - \bm p}} = \hat{\bm{k}} \cdot \widehat{\bm{k}-\bm{p}}$.
Then, the dimensionless power spectrum of tensor modes is defined as
\begin{align}
\langle \hat{h}^s_{\bm{k}} \hat{h}^{s'}_{\bm{k}'} \rangle &= \langle \hat{h}^s_{\bm{k},\rm v} \hat{h}^{s'}_{\bm{k}',\rm v} \rangle + \langle \hat{h}^s_{\bm{k},\rm s} \hat{h}^{s'}_{\bm{k}',\rm s} \rangle \notag \\
&\equiv (2\pi)^3\delta^{ss'}\delta(\bm{k} + \bm{k}')\dfrac{2\pi^2}{k^3}\left(\mathcal{P}_{h,\rm v}(k) + \mathcal{P}^{ss}_{h, \rm s}(k) \right) \ ,
\end{align}
where the cross terms vanish for linear perturbations.
One can find
\begin{align}
\mathcal{P}^{++}_{h, \rm s}(k)|_{\tau} &= \dfrac{8k^3}{\pi^2H^4\Mpl^4}\int\dfrac{d\bm{p}}{(2\pi)^3}\mathcal{F}^{++}(\hat{\bm{p}}, \ \widehat{\bm k - \bm p})\left|\int_{\tau_{\rm min}}^\tau \dfrac{d\tau'}{\tau'} \dfrac{x^3 - y^3}{3y^3}E_pE_{|\bm k -\bm p|}\right|^2 \ , \label{Phs++} \\
\mathcal{P}^{\times\times}_{h, \rm s}(k)|_{\tau} &= \dfrac{8k^3}{\pi^2H^4\Mpl^4}\int\dfrac{d\bm{p}}{(2\pi)^3}\mathcal{F}^{\times\times}(\hat{\bm{p}}, \ \widehat{\bm k - \bm p})\left|\int_{\tau_{\rm min}}^\tau \dfrac{d\tau'}{\tau'} \dfrac{x^3 - y^3}{3y^3}E_p E_{|\bm k -\bm p|}\right|^2 \ ,
\label{Phsxx}
\end{align}
where
\begin{align}
\mathcal{F}^{++} &\equiv \cos^2(2\phi_{\hat{\bm{p}}})\left( \cos^2\theta_{\hat{\bm{p}}}\cos^2\theta_{\widehat{\bm k - \bm p}} + 1 \right) + \sin^2(2\phi_{\hat{\bm{p}}})\left( \cos^2\theta_{\hat{\bm p}} + \cos^2\theta_{\widehat{\bm k - \bm p}} \right) \ , \\
\mathcal{F}^{\times\times} &\equiv \sin^2(2\phi_{\hat{\bm{p}}})\left( \cos^2\theta_{\hat{\bm{p}}}\cos^2\theta_{\widehat{\bm k - \bm p}} + 1 \right) + \cos^2(2\phi_{\hat{\bm{p}}})\left( \cos^2\theta_{\hat{\bm p}} + \cos^2\theta_{\widehat{\bm k - \bm p}} \right) \ ,
\end{align}
and we have approximated the lower bound of the time integral by $|\tau_{\text{min}}| = \text{min}(1/p, 1/|\bm{k}-\bm{p}|)$, because the dark photon can grow only after the horizon crossing and we focus on the contribution from the super-horizon modes. 

For our convenience, we perform the numerical evaluation of
the time integral and that of the momentum integral separately.
To do this, we define the following Gaussian fitting function for the electric component:
\begin{equation}
E_k(x) \simeq E_{k,\rm fit}(x) \equiv \dfrac{H^2}{\sqrt{2k^3}}A_{\text{peak}}(k)\exp\left[ -\dfrac{1}{2}\left.\dfrac{d n}{dN}\right|_{N=N_2}(\ln(\tau/\tau_{2}))^2 \right] \ , \label{eq: fit}
\end{equation}
where $A_{\rm peak}(k)$ is the maximum amplitude and $\tau_2$ is the conformal time at which $N = N_2$. 
Plugging Eq.~\eqref{eq: fit} into Eqs.~\eqref{Phs++} and \eqref{Phsxx},
at the super-horizon limit $\tau \rightarrow 0$,
i.e.~$x \ll y \ll 1$,
we find
\begin{align}
&\mathcal{P}^{++}_{h, \rm s}(k)|_{\tau\rightarrow0} = \mathcal{P}^{\times\times}_{h, \rm s}(k)|_{\tau\rightarrow0} \notag \\
&\simeq \dfrac{1}{\pi^2} \dfrac{\mathcal{F}^2H^4}{9\Mpl^4}\int\dfrac{dp_*d\cos\theta_{\hat{\bm{p}}}}{(2\pi)^2}\left(1 + \cos^2\theta_{\hat{\bm p}}\right)\left(1 + \cos^2\theta_{\widehat{\bm k - \bm p}} \right)\dfrac{A^2_{\rm peak}(p)A^2_{\rm peak}(|\bm{k}-\bm{p}|)}{p_* \, \vert \bm k - \bm p \vert_*^{3}} \ ,
\end{align}
where $p_* \equiv p/k \ , \ |\bm k - \bm p|_{*} \equiv |\bm k - \bm p|/k$ and $\mathcal{F}$ is the numerical factor obtained by the time integration,
\begin{equation}
\mathcal{F} \equiv \int_{-\infty}^\infty \frac{d \tau'}{\tau'} \, \exp\left[ -\left.\dfrac{d n}{dN}\right|_{N=N_m}(\ln(\tau'/\tau_{m}))^2 \right] = \sqrt{\pi}\left(\left.\dfrac{d n}{dN}\right|_{N=N_m}\right)^{-1/2} \label{eq: F} \ .
\end{equation}
Here, $N_m (\tau_m)$ is the time when the amplitude of the electric component is maximized.
Note that the time integration can be extended to $\pm \infty$ 
($-\infty \to 0$ in the terms of $\tau'$)
since the integral has its support almost around the peak of the electric mode function.
The cross power spectrum $\mathcal{P}^{+\times/\times+}_{h, \rm s}(k)$
vanishes due to the cancellation in integrating the periodic function $\sin(2\phi_{\hat{\bm{p}}})\cos(2\phi_{\hat{\bm{p}}})$.

\begin{figure}[tbp]
\center
\includegraphics[width=0.8\textwidth]{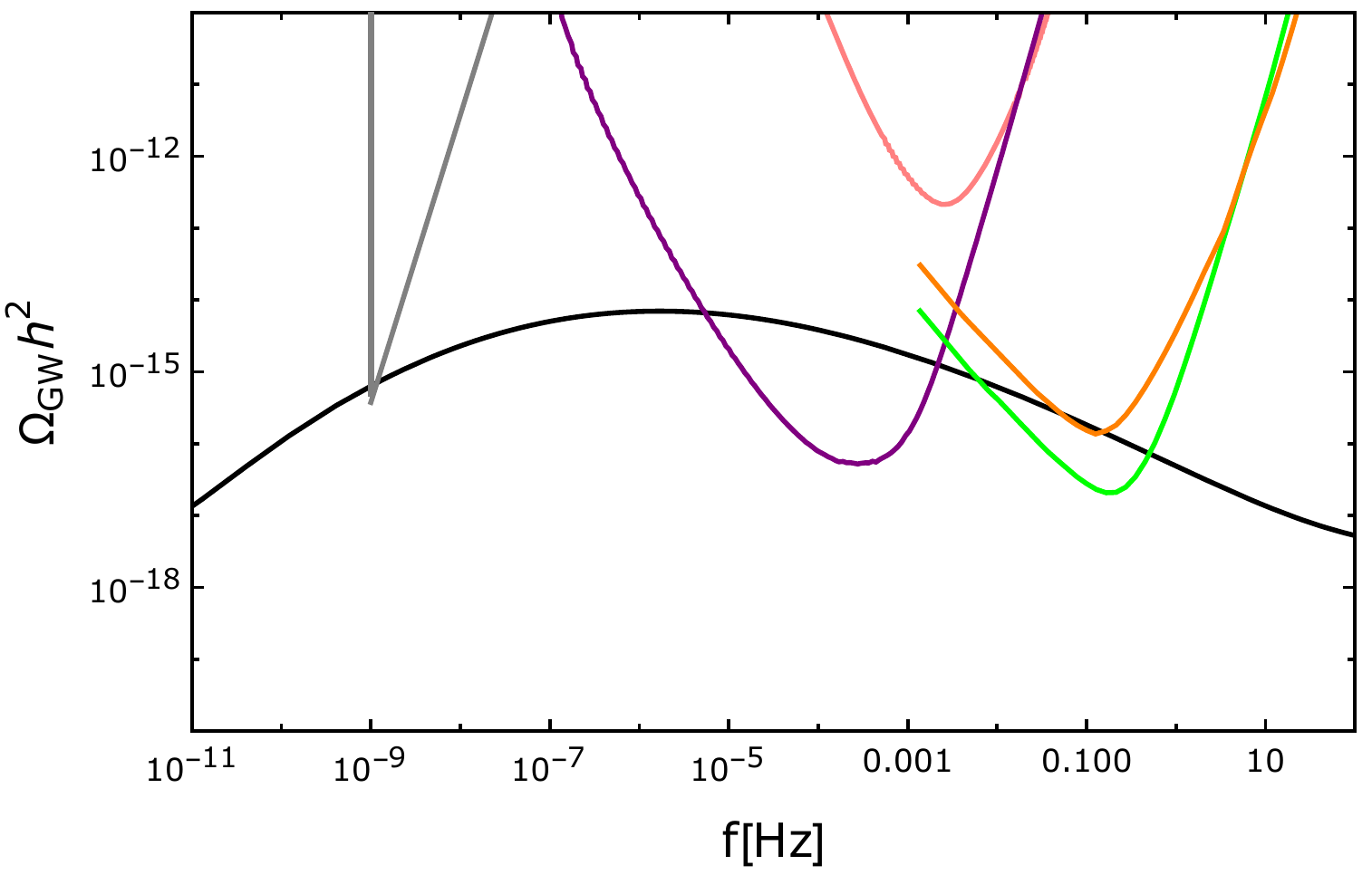}
\caption{The GW spectrum sourced by the dark photon field during inflation.
The sensitivity curves of SKA \cite{Kramer:2004hd} (gray), LISA \cite{Smith:2019wny} (pink), DECIGO \cite{Braglia:2021fxn} (orange), BBO \cite{Braglia:2021fxn} (green) and $\mu$Ares \cite{Sesana:2019vho} (purple)
are also shown. We here take $R = 10^{-2}$, $\Lambda_1 = 5\times10^{-3}\Mpl$, $r_{\rm v} = 5\times10^{-4}$, $\tilde{\sigma}_i = 1.2525$, $c = 7.525\times10^{-3}$, which correspond to $m_{\gamma'} \simeq 1.64\times10^{-13} \ \text{eV}$.
}
\label{fig: GWPlot}
\end{figure}

Let us now evaluate the logarithmic energy density of GWs at present,
\begin{equation}
\Omega_{\rm GW}(k) \equiv \dfrac{1}{\rho_c}\dfrac{d\rho_{\rm GW}}{d\ln k} \ ,
\end{equation}
where $\rho_c = 3\Mpl^2H_0^2$ is the present critical energy density of the Universe.
Using the entropy conservation law, $\Omega_{\rm GW}$ is related to the power spectrum of primordial tensor modes as \cite{Smith:2005mm},%
\footnote{The factor $4$ difference in the numerical factor compared to the expression in \cite{Smith:2005mm} comes from the fact that our definition of $h_{ij}$ contains $1/2$, as seen right above \eqref{eq:eom_tensor}, which amounts to $1/4$ in terms of $\mathcal{P}_h$.}
\begin{equation}
\Omega_{\rm GW}(k)h^2 \simeq 
6.85 \times 10^{-7}
\left(\dfrac{g_*}{100}\right)^{-1/3}
\left(\mathcal{P}_{h,\rm v}(k) + \mathcal{P}^{++}_{h, \rm s}(k)
+ \mathcal{P}^{\times \times}_{h, \rm s}(k) \right) \ .
\end{equation}
Figure~\ref{fig: GWPlot} depicts the power spectrum of the GW energy density.
On large scales, the contribution from the vacuum modes is dominant and the spectrum is scale-invariant.
At intermediate scales, however, its magnitude gets amplified by the contribution of the sourced tensor modes and has a peak at a frequency around $\mu$Hz.
This spectral shape is determined by the background time evolution of the index $n$.
Even if the vacuum tensor-to-scalar ratio is small, the sourced power spectrum is potentially testable with the next generation pulsar timing array measurement (SKA) and the projected space-based laser interferometers such as DECIGO, BBO and $\mu$Ares.
On the other hand, it would be challenging to test it by LISA.

%
%

\section{Conclusion}\label{conclusion}

We have explored a mechanism to produce a light dark photon DM through a coupling between
the dark photon field and a spectator scalar field that does not play a role in
the inflationary expansion of the Universe but is rolling down its potential during the inflation.
The motion of the spectator field efficiently produces dark photons with large wavelengths
which become non-relativistic before the time of matter-radiation equality.
We have constructed the mechanism such that the spectrum of wavelengths is peaky
so that the constraint from the isocurvature perturbation can be evaded.
The correct relic abundance is then achieved over a wide range of the dark photon mass.
Depending on the model parameter set, we have found that our result could provide the dark photon DM with mass down to $m_{\gamma'} 
\approx 10^{-13} \ \text{eV}$.
Our mechanism favors high-scale inflation models which can be tested in future observations.
Furthermore, fluctuations of the dark photon field during inflation produce GWs
detectable at future space-based interferometers.

It would be interesting to consider the evolution of the dark photon energy density after inflation.
At the end of inflation, the speed of the spectator field increases and the provided dark photon may backreact to the evolution.
This effect would more or less modify the allowed parameter region of the dark photon DM.
Our result is sensitive to both the amount of the dark photon production and the reheating temperature, and therefore, if there were to be an efficient enhancement around or after the end of inflation, a mass of the dark photon DM that is smaller than the one in our current study may be within the reach. This, however, likely requires non-linear studies of the coupled system between the spectator scalar and the dark photon and is beyond our current scope.

Regarding to the generation of the dark photon, we have assumed no homogeneous component of the dark photon because our scenario does not generate the dark photon on large scales.
In this sense, the resultant power spectrum is statistically isotropic on large scales.
However, it might be worthwhile to prove the statistical anisotropy of tensor modes on intermediate scales,
which would serve as another venue of interesting signatures of the primordial Universe.
We leave these issues to a future study.

\section*{Acknowledgements}

We would like to thank Eiichiro Komatsu for fruitful discussions.
YN is supported by Natural Science Foundation of China under grant No.~12150610465.
RN is in part supported by RIKEN Incentive Research Project grant.
IO acknowledges the support from JSPS Overseas Research Fellowship and JSPS KAKENHI Grant No.~JP20H05859 and 19K14702.

\appendix

\section{Dark photon production with constant $n$}
\label{app:constantn}

In this appendix, we briefly summarize the calculation of dark photon production for constant $n$, denoting it by $n_0$.
The existing calculations in the literature can be found in, e.g., \cite{Bamba:2003av,Barnaby:2012tk,Nakai:2020cfw}.
In this case, $I$ simply behaves as a function of the scale factor, $I \propto a^{n_0}$, and Eq.~\eqref{eq: Ve1} is reduced to
\begin{equation}
\partial_\tau^2V_k + \left[ k^2 - \dfrac{n_0 (n_0 +1)}{\tau^2} + \dfrac{a^2m_{\gamma'}^2}{I^2} \right] V_k = 0 \ . \label{eq: Ve2}
\end{equation}
This equation is derived under the de-Sitter approximation, i.e.~$a \simeq -1 / (H \tau)$.
If the third term in the square parentheses of the above equation is negligible in comparison with the other terms,
an instability could occur on the super-horizon regime 
$-k\tau \le \sqrt{n_0(n_0+1)}$
for a certain range of $n_0$.
To realize this instability, we consider a negative branch of the index $n_0 < 0$, where $I$ becomes a decreasing function in time \cite{Nakai:2020cfw}.
\footnote{This branch is also appropriate for avoiding a strong coupling problem of the dark photon interacting with other matter sectors during inflation \cite{Demozzi:2009fu}.
As can be seen in the action, normalization of $I$ is relevant for the magnitude of an effective coupling strength 
because it is inversely proportional to $I$.
Since $I$ is exponentially large
at early times,
the coupling strength is highly suppressed throughout the inflationary period.
This can be observed already from the effective mass term, which is essentially an illustration of the coupling to matter, as $m_{\gamma'}/I$ is suppressed earlier during inflation for $n_0 <0$ and would be divergent if $n_0>0$.}
The condition of the instability holds for a whole period of inflation if
\begin{equation}
\dfrac{m_{\gamma'}^2}{I(t_{\rm end})^2H^2} \ll n_0(n_0+1)
\end{equation}
is satisfied at the end of inflation $t = t_{\rm end}$.
Then, 
neglecting the mass term, we obtain a solution of $V_k$ with the Bunch-Davies initial condition,
\begin{equation}
V_k = \frac{i}{\sqrt{2k}} \sqrt{\frac{- \pi k \tau}{2}} \, H^{(1)}_{-n_0 - 1/2} (-k \tau) \; ,
\label{eq: Hankel}
\end{equation}
given by the Hankel function of the first kind.
We have chosen the arbitrary initial phase such that the mode function becomes real at the leading order outside the horizon $- k \tau \ll 1$.
The mode functions of the dark electric and magnetic fields, defined in \eqref{eq:EkBk}, are then given by
\begin{align}
E_k &\equiv - \dfrac{I}{a^2}\dfrac{d}{d \tau}\left(\dfrac{V_k}{I}\right) = 
\frac{- i \sqrt{\pi}}{2 k^{3/2}} \, H^2 \left( - k \tau \right)^{5/2} H^{(1)}_{-n_0+1/2}(-k\tau) \ , 
\label{eq: eHankel} \\
B_k &\equiv \frac{k}{a^2} \, V_k = \frac{i \sqrt{\pi}}{2 k^{3/2}} \, H^2 \left( - k \tau \right)^{5/2} H^{(1)}_{-n_0-1/2}(-k\tau)
\label{eq: mHankel} \ .
\end{align}
Using the asymptotic form of the Hankel function,
\begin{equation}
H^{(1)}_{\nu} (x) \simeq - i \, \frac{\Gamma(\nu)}{\pi} \left( \frac{2}{x} \right)^{\nu} \; , \qquad ( x \rightarrow 0 \, , \; {\rm Re} \, (\nu) >0) \; ,
\end{equation}
the expressions \eqref{eq: eHankel} and \eqref{eq: mHankel} in the super-horizon limit $(|k\tau| \rightarrow 0)$ are given by
\begin{align}
E_k &\simeq \frac{- \Gamma\big( \frac{1}{2} - n_0 \big)}{2^{n_0 + 1/2} \sqrt{\pi} \, k^{3/2}} \, \frac{H^2}{( - k \tau )^{-n_0 -2}} \; , \qquad \left( n_0 < \frac{1}{2} \right) \; ,
\label{eq: eHankels} \\
B_k &\simeq \frac{\Gamma\big( -\frac{1}{2} - n_0 \big)}{2^{n_0 + 3/2} \sqrt{\pi} \, k^{3/2}} \, \frac{H^2}{( - k \tau )^{- n_0 -3}} \; , \qquad \left( n_0 < - \frac{1}{2} \right) \; .
\label{eq: mHankels}
\end{align}
Noting $-k\tau = e^{N_k-N} \ll 1$, where $N_k \equiv \ln(k/H)$ is the number of e-foldings at which fluctuations with momentum mode $k$ exits the horizon, their amplitudes are proportional to
\begin{align}
|E_k| &\propto e^{(-n_0-2)(N - N_k)} \ , \quad 
|B_k| \propto e^{(-n_0-3)(N - N_k)} \, , \qquad  (N - N_k > 0)
\label{eq: emsuperhorizon} \ .
\end{align}
Therefore, on super-horizon scales the electric field is proportional to $a^{-(n_0+2)}$ and grows when $n_0 < -2$ is satisfied.
On the other hand, the magnetic field evolves as $a^{-(n_0+3)}$
and grows for $n_0 < -3$.
In the parameter domain $n_0 < 0$ of our interest, it is hence clear that (i) sufficient production of the dark photon energy is achieved for $n_0 < -2$, and (ii) the dark electric field dominates over the magnetic counterpart.

When $n$ is not constant but dynamically evolves in time, $n \neq n_0$,
the solution \eqref{eq: Hankel} for a constant value of $n_0$ does not necessarily hold.
However, particle production still occurs when $|n|$ becomes greater than a threshold value $|n|=2$,
provided the time variation of $n$ is sufficiently small.
As we explore in the main text, this fact enables to validate a scenario of an evolving $n$ to realize
production of the dark photon on scales much smaller than the CMB scale.

For completeness, we next consider the evolution of the longitudinal mode.
It is characterized by the time variation of the function $\partial_\tau^2z_k/z_k$ which is given by
\begin{equation}
\dfrac{\partial_\tau^2 z_k}{z_k} = \dfrac{1}{\tau^2} \, \dfrac{2p^4 - (2n^2 - 7n + 1 + \tau \, dn/d\tau)p^2M^2 + (n^2 + n - \tau \, dn/d\tau)M^4}{(p^2+M^2)^2} \ ,
\end{equation}
where we have defined $p \equiv k/a$ and $M \equiv m_{\gamma'}/I$.
With a constant $n \simeq n_0.$ and a negligible mass in Eq.~\eqref{eq: Xe1},
we obtain
\begin{align}
&\partial_\tau^2 X_k + \left(k^2 - \dfrac{2}{\tau^2}\right)X_k \simeq 0 \qquad (p \gg M) \ , 
\label{eq:eom_long_largep}\\
&\partial_\tau^2 X_k - \dfrac{n_0 (n_0 + 1)}{\tau^2}X_k \simeq 0 \qquad (p \ll M) \ .
\label{eq:eom_long_smallp}
\end{align}
Then, for the Bunch-Davies initial condition, we find
\begin{align}
X_k &\simeq \dfrac{e^{-ik\tau}}{\sqrt{2k}}\left( 1 - \dfrac{i}{k\tau} \right) \qquad (p \gg M) \ , \label{eq: Xsol} \\
X_k &\simeq C_1(-\tau)^{n_0+1} + \dfrac{C_2}{(-\tau)^{n_0}} \qquad (p \ll M) \ ,
\end{align}
where $C_{1,2}$ are integration constants determined by connecting the two solutions at the conformal time $\tau_{\text{NR},k}$ when $p = M$ \cite{Nakai:2020cfw}: 
\begin{equation}
C_1 = \sqrt{\dfrac{k}{2}}
\dfrac{(-\tau_{\text{NR},k})^{-n_0}}{2n_0+1} \ , \qquad C_2 = \dfrac{(-\tau_{\text{NR},k})^{n_0-1}}{\sqrt{2}k^{3/2}} \ .
\end{equation}
In order to obtain $C_1$ and $C_2$, we have assumed (i) $-k \tau_{\rm NR} \ll 1$, and (ii) the continuity of the energy density associated with the longitudinal mode, not the continuity of $X_k$ and its time derivative.
Therefore, in either case of $p \ll M$ and $p \gg M$, the contribution from the longitudinal mode is subdominant in comparison to the exponentially enhanced energy density of the transverse modes for $n < -2$, and we neglect the former in what follows.

\bibliographystyle{apsrev4-1}
\bibliography{Ref.bib}

\end{document}